\documentclass[11pt,reqno]{article}

\include{psfig}
\include{epsf}

\usepackage{color}
\usepackage{amssymb} 

\newtheorem{theorem}{Theorem}[section]

\newtheorem{proposition}[theorem]{Proposition}

\newtheorem{definition}{Definition}[section]

\newtheorem{simulation algorithm}{Simulation Procedure}[section]

\catcode`\@=11
\renewcommand{\section}{
         \setcounter{equation}{0}
         \@startsection {section}{1}{\z@}{-3.5ex plus -1ex minus
         -.2ex}{2.3ex plus .2ex}{\normalsize\bf}
}
\renewcommand{\subsection}{
         \@startsection {subsection}{1}{\z@}{-3.5ex plus -1ex minus
         -.2ex}{2.3ex plus .2ex}{\normalsize\bf}
}
\catcode`\@=12

\def\reals{{\rm\vrule depth0ex width.4pt\kern-.08em R}}
\def\bbbz{{\mathchoice {\hbox{$\sf\textstyle Z\kern-0.4em Z$}}
{\hbox{$\sf\textstyle Z\kern-0.4em Z$}}
{\hbox{$\sf\scriptstyle Z\kern-0.3em Z$}}
{\hbox{$\sf\scriptscriptstyle Z\kern-0.2em Z$}}}}

\newcommand{\nc}{\newcommand}
\nc{\W}{{\bf W}}
\nc{\A}{{\bf A}}
\nc{\bL}{{\bf L}}
\nc{\bH}{{\bf H}}
\nc{\C}{{\cal C}}

\def\eq#1{(\ref{e:#1})}

\def\elabel#1{\label{e:#1}}

\begin{document}
\begin{center}
\Large\bf $n$-Qubit Operations on Sphere and Queueing Scaling Limits for Programmable Quantum Computer
\end{center}
\begin{center}
\large\bf Wanyang Dai~\footnote{The project is funded by National Natural Science Foundation
of China with Grant No. 11771006, Grant No. 10971249, and Grant No. 11371010.}
\end{center}
\begin{center}
\small Department of Mathematics\\
and State Key Laboratory of Novel Software Technology\\
Nanjing University, Nanjing 210093, China\\
Email: nan5lu8@nju.edu.cn\\
September 29, 2021
\end{center}

\vskip 0.1 in
\begin{abstract}

We study $n$-qubit operation rules on $(n+1)$-sphere with the target to help developing a (photon or other technique) based programmable quantum computer. In the meanwhile, we derive the scaling limits (called reflecting Gaussian random fields on a $(n+1)$-sphere) for $n$-qubit quantum computer based queueing systems under two different heavy traffic regimes. The queueing systems are with multiple classes of users and batch quantum random walks over the $(n+1)$-sphere as arrival inputs. In the first regime, the qubit number $n$ is fixed and the scaling is in terms of both time and space. Under this regime, performance modeling during deriving the scaling limit in terms of balancing the arrival and service rates under first-in first-out and work conserving service policy is conducted. In the second regime, besides the time and space scaling parameters, the qubit number $n$ itself is also considered as a varying scaling parameter with the additional aim to find a suitable number of qubits for the design of a quantum computer. This regime is in contrast to the well-known Halfin-Whitt regime.\\

\end{abstract}

\section{Introduction}

As pointed out by U.S. Los Angeles based Six Industrial Revolution Forum (SIR Forum~\cite{sirfor}), quantum computer based quantum computing will be the core technology in the future industrial revolution. It will provide the required super-computing power for the quickly developing information communication system, big data service, digital economy, blockchain, and even the future Internet of quantum blockchains (see, e.g., Arule {\it et al.}~\cite{aru:quasup}, Dai~\cite{dai:plamod,dai:quacom}, Deutsch~\cite{deu:quacom}, Feynman~\cite{fey:quamec}, Harrow and Montannaro~\cite{harmon:quacom}, Luo {\it et al.}~\cite{lo:quatel}, Nielsen and Chuang~\cite{niechu:quacom}, Rajan and Visser~\cite{rajvis:quablo}, Zhong {\it et al.}~\cite{zho:quacom}).

However, to make a quantum computer programable for the purpose to interact with real-world applications, it should have buffer storage and data read/write capability in addition to its processing capability (see, e.g., the illustration in Figure~\ref{quantumchannel}),
\begin{figure}[tbh]
\centerline{\epsfxsize=5.5in\epsfbox{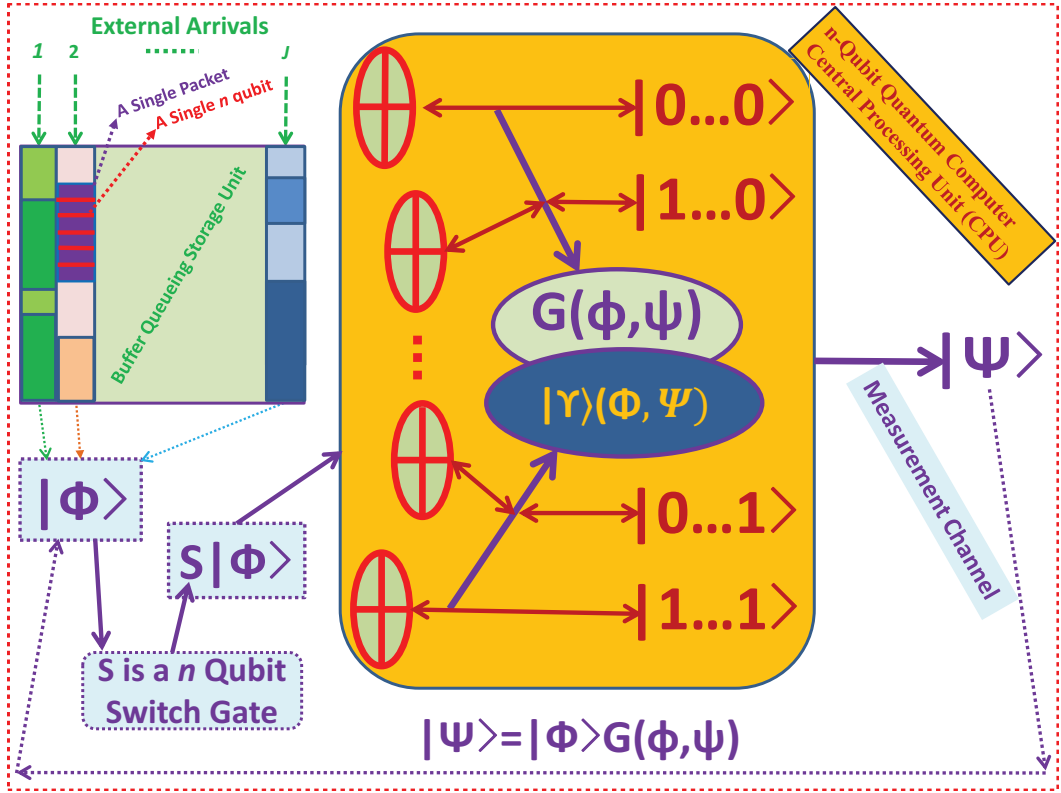}}
\caption{\footnotesize A quantum buffer queueing storage, quantum computer central processing unit (CPU) with $n$-qubit operations $|\Upsilon\rangle(\Phi,\Psi)$ over unit $(n+1)$-sphere as introduced in \eq{jkdsalp}, and quantum channel measurement interactive system with channel function $G(\Phi,\Psi)$ derived in \eq{channelphipsi}.}
\label{quantumchannel}
\end{figure}
which falls into the research scope concerning a quantum queueing system for its internal qubit data flow modeling, management, and related performance analysis (see, e.g., Dai~\cite{dai:plamod,dai:quacom}, Gawron {\it et al.}~\cite{gawkur:modqua}, and Mandayam {\it et al.}~\cite{manjag:clacap}). Nevertheless, since this area is just getting started, a broad view concerning the system framework and research methodology needs to be put forward. Therefore, in this paper, we make such an attempt.

More precisely, we study $n$-qubit operation rules on $(n+1)$-sphere concerning addition (+), substraction (-), multiplication ($\ast$), and division ($\slash$) with the target to help developing a (photon or other technique) based programmable quantum computer. In the meanwhile, we derive the scaling limits (called reflecting Gaussian random fields (RGRFs) on a $(n+1)$-sphere (denoted by $S^{n+1}$)) for $n$-qubit quantum computer based queueing systems under two different heavy traffic regimes. The queueing systems are with multiple classes of users and batch quantum random walks over the $(n+1)$-sphere as arrival inputs. In the first heavy traffic regime that corresponds to the way used in Dai~\cite{dai:broapp,{dai:optrat}} and Dai and Dai~\cite{daidai:healim}, the qubit number $n$ is fixed and the scaling is with respect to both time and space. Under this regime, we are interested in the performance modeling through deriving RGRF in terms of reasonably balancing the arrival and service rates under first-in first-out and work conserving service policy. In the second heavy traffic regime, besides the time and space scaling parameters, the qubit number $n$ itself is also considered as a varying scaling parameter with the additional aim to find a suitable number of qubits for the design of a quantum computer. The second regime is in contrast to the well-known Halfin-Whitt regime in~\cite{halwhi:heatra} where the number of servers is considered as a scaling parameter.

\begin{figure}[tbh]
\centerline{\epsfxsize=5.5in\epsfbox{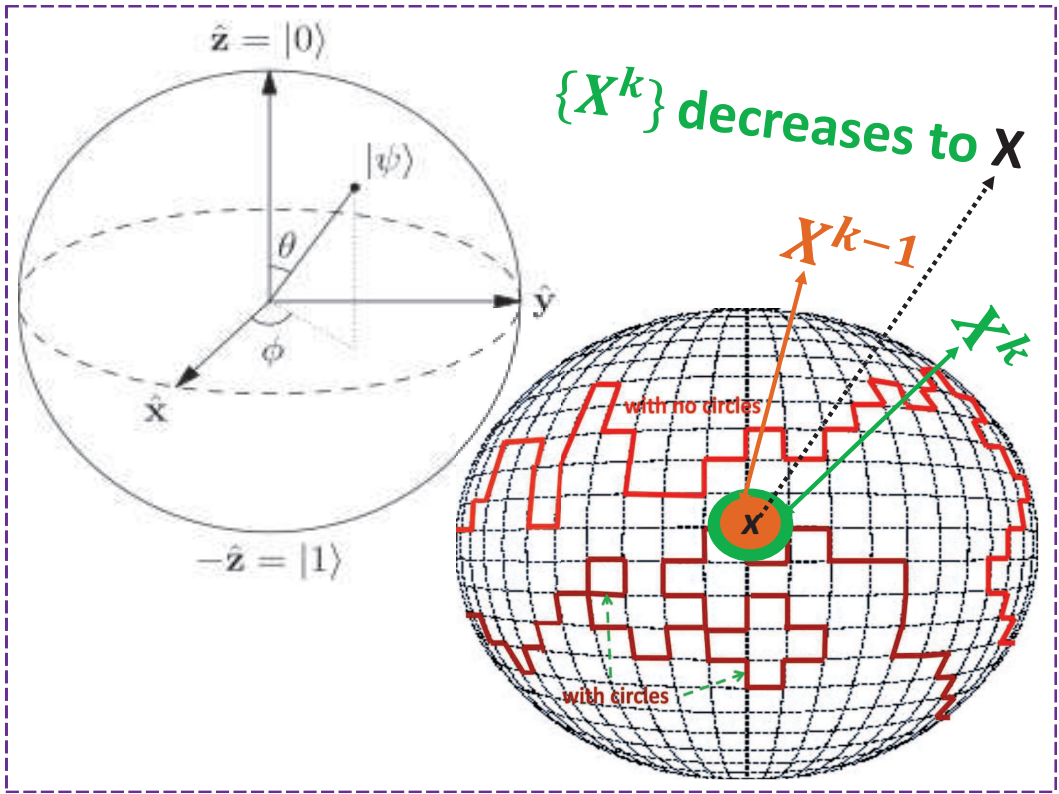}}
\caption{\footnotesize Single qubit representation, quantum random walk on $S^{2}$, and converging sets.}
\label{qubitrandomwalk}
\end{figure}

Note that, as an illustration of quantum random walk, a single qubit with $n=1$ is used to denote a particle spinning up and down at the same time. In this case, a pure qubit state $|\Psi\rangle$ can be denoted by any point on the $2$-sphere $S^{2}$ (called Bloch sphere, the surface of a ball) with corresponding angles $\theta$ and $\phi$ as shown in the upper-left graph of Figure~\ref{qubitrandomwalk},
i.e., $|\Psi\rangle=\psi_{0}|0\rangle+\psi_{1}|1\rangle\;\;\mbox{with}\;\;\psi_{0}=\mbox{cos}\left(\theta/2\right)\;\;\mbox{and}\;\;
\beta=e^{i\phi}\mbox{sin}(\theta/2)$. Thus, the red and brown curves in the lower-right graph of Figure~\ref{qubitrandomwalk}
can be used to represent two sample paths of a single-qubit quantum random walk over $S^{2}$ and readers are referred to Kempe~\cite{kem:quaran} and Ko$\check{s}\acute{\imath}$k~\cite{kosbuz:quawal} for the concept concerning a quantum random walk.

The derived scaling limits over a set ${\cal M}$ where ${\cal M}=S^{n+1}$ or ${\cal M}=S^{\infty}$ (the limit of $S^{n+1}$ as $n$ tends to infinity) can be presented in the following definition.
\begin{definition}\label{defrgrf}
A $(1+n)$-parameter time-space random field $\tilde{V}(t,x)\in R_{+}$ with $(t,x)\in [0,\infty)\times{\cal M}$ is called a RGRF over a set ${\cal M}$ if it has the form
\begin{eqnarray}
&&\tilde{V}(t,x)=\mu(t,x)+\tilde{U}(t,x)+\tilde{I}(t,x),
\elabel{rgrfp1}
\end{eqnarray}
where, $\mu(t,x)$, $\tilde{U}(t,x)$, and $\tilde{I}(t,x)\in D[0,\infty)$ for each given $x\in{\cal M}$. Furthermore, $\tilde{I}(t,x)$ is nondecreasing in $t\in[0,\infty)$ for each fixed $x\in{\cal M}$ and it can increase only at a time $t$ when $\tilde{V}(t,x)=0$, i.e.,
\begin{eqnarray}
&&\int_{0}^{\infty}\tilde{V}(t,x)d\tilde{I}(t,x)=0\;\;\;\mbox{for each fixed}\;\;\;x\in{\cal M}.
\elabel{rgrfp2}
\end{eqnarray}
\end{definition}
Note that, in the definition, $D[0,\infty)$ denotes the well-known single-dimensional Skorohod space of all right-continuous functions with left-limits (see, e.g., Ethier and Kurtz~\cite{ethkur:marpro}).

The comparisons between our current research and the existing ones on quantum queueing systems can be summarized as follows. In the study of Gawron {\it et al.}~\cite{gawkur:modqua}, the authors present a quantum queueing model via the method of discrete time Markov chain. Nevertheless, in our current study, our queueing model is related to batch quantum random walks with multiclass service requirements and hence general doubly stochastic renewal reward random fields (DSRRRFs) over $S^{n}$ are involved. Thus, different heavy traffic regimes are introduced to our discussion. In the study of Dai~\cite{dai:plamod,dai:quacom}, the quantum computers are assumed to have already been built and we try to use them to propose a further quantum computer based network (or a quantum computer based quantum-cloud computing system). However, since quantum computers are still under developments, thus as an initiative presented in this paper, we turn to more deep study concerning how to design and build quantum computers themselves, which includes establishing our new $n$-qubit operation rules on $(n+1)$-sphere. In the study of Mandayam {\it et al.}~\cite{manjag:clacap}, the authors derive the classic capacity of additive quantum queue-channels, which has different system formulation and purpose from our current research.

Finally, the rest of the paper is organized as follows. In Section~\ref{lpqwasw}, we propose our $n$-qubit operation rules through stating a proposition with illustration and examples. In Section~\ref{queuelimit}, we establish our performance models though queueing scaling limits by proving heavy traffic limit theorems. In Section~\ref{fsgdhja} (Appendix of our crrent paper), we justify our designed $n$-qubit operation rules by proving our stated proposition in Section~\ref{lpqwasw}. In Section~\ref{cremark}, we conclude the current research with remarks.

\vskip 1cm
\section{$n$-Qubit Operations on $(n+1)$-Sphere}\label{lpqwasw}

As claimed in the introduction of this paper, a programmable quantum computer should consist of buffer queueing storage unit, central processing unit (CPU), and quantum channel measurement capability as explained in Figure~\ref{quantumchannel}. In this section, we study $n$-qubit operations addition (+), substraction (-), multiplication ($\ast$), and division ($\slash$) within the designed CPU as shown in Figure~\ref{quantumchannel}. Illustration with examples and the channel measurement formula $G(\Phi,\Psi)$ (proposed in Figure~\ref{quantumchannel}) are also presented.

\vskip 1cm
\subsection{The Operations}\label{alhjsk}

In a quantum computer system, the basic information unit is a $n$-qubit with $n\in\{1,2,...\}$ and can be expressed through the conventional complex column-vector oriented ket-notation. More precisely,  a state $|\Psi\rangle$ of $n$-qubit register is represented by
\begin{eqnarray}
|\Psi\rangle=\sum_{j_{l}\in\{0,1\},\;l\in\{1,...,n\}}\psi_{j_{1}...j_{n}}|j_{1}...j_{n}\rangle,
\elabel{nqubi}
\end{eqnarray}
where, $|j_{1}...j_{n}\rangle$ for each $j_{l}\in\{0,1\}$ and $l\in\{1,...,n\}\}$ is called an eigenstate and there are total number $2^{n}$ of those eigenstates. The basis of bit strings $\{j_{1}...j_{n}:\;j_{l}\in\{0,1\},\;l\in\{1,...,n\}\}$ is the computational basis with the associated complex coefficients represented by $\{\psi_{j_{1}...j_{n}}\}$.
The summation of the squares of the coefficients' absolute values in \eq{nqubi} must satisfy
\begin{eqnarray}
\sum_{j_{l}\in\{0,1\},\;l\in\{1,...,n\}}\left|\psi_{j_{1}...j_{n}}\right|^{2}=1.
\elabel{cnqubi}
\end{eqnarray}
For a bit string $j_{1}...j_{n}$, the value $|\psi_{j_{1}...j_{n}}|^{2}$ denotes the probability of the system that is found in the $(j_{1}...j_{n})^{th}$ state after a measurement. Nevertheless, since a complex number encodes not just a magnitude but also a direction in the complex plane, the phase difference between any two coefficients is a valuable parameter and represents a key difference between quantum computing and probabilistic traditional computing. Under this computational basis, a state $|\Psi\rangle$ of $n$-qubit register can be represented by its coefficients $\{\psi_{j_{1}...j_{n}}\}$. More precisely, for each index $j_{1}\cdot\cdot\cdot j_{n}$ in \eq{nqubi} with associated integer $j_{l}\in\{0,1\}$ and each number $l\in\{1,...,n\}$, we re-index it through an index $h$ expressed by
\begin{eqnarray}
h\;=\;
2^{n-1}j_{n}+2^{n-2}j_{n-1}+...+2j_{2}+j_{1}.
\elabel{reindexi}
\end{eqnarray}
Then, we can rearrange the coefficients $\{\psi_{j_{1}...j_{n}}\}$ through the index $h$ as $\{\psi_{h+1}\}$, where,
\begin{eqnarray}
h\in\{h_{0},h_{1},...,h_{2^{n}-1}\}=\{0,1,...,2^{n}-1\}.
\nonumber
\end{eqnarray}
Hence, we can present all the coefficients $\{\psi_{h+1}\}$ in terms of a corresponding spherical coordinate $\theta=(\theta_{1},\cdot\cdot\cdot,\theta_{2^{n}})$ as follows,
\begin{eqnarray}
\left\{\begin{array}{ll}
\psi_{1}&=\;\;\;\;\;\cos(\theta_{1}),\\
\psi_{2}&=\;\;\;\;\;\sin(\theta_{1})\cos(\theta_{2}),\\
\psi_{3}&=\;\;\;\;\;\sin(\theta_{1})\sin(\theta_{2})\cos(\theta_{3}),\\
     &\cdot\;\\
     &\cdot\;\\
     &\cdot\;\\
\psi_{2^{n}-1}&=\;\;\;\;\;\sin(\theta_{1})\sin(\theta_{2})\cdot\cdot\cdot\sin(\theta_{2^{n}-2})\cos(\theta_{2^{n}-1}),\\
\psi_{2^{n}}&=\;\;\;\;\;e^{i\theta_{2^{n}}}\sin(\theta_{1})\sin(\theta_{2})\cdot\cdot\cdot\sin(\theta_{2^{n}-1}),
\end{array}
\right.
\elabel{sphericalco}
\end{eqnarray}
where, $\theta_{i}\in[0,\pi/2]$ for each $i\in\{1,...,2^{n}-1\}$ and $\theta_{2^{n}}=\varphi\in[0,2\pi]$.

\begin{figure}[tbh]
\centerline{\epsfxsize=5.5in\epsfbox{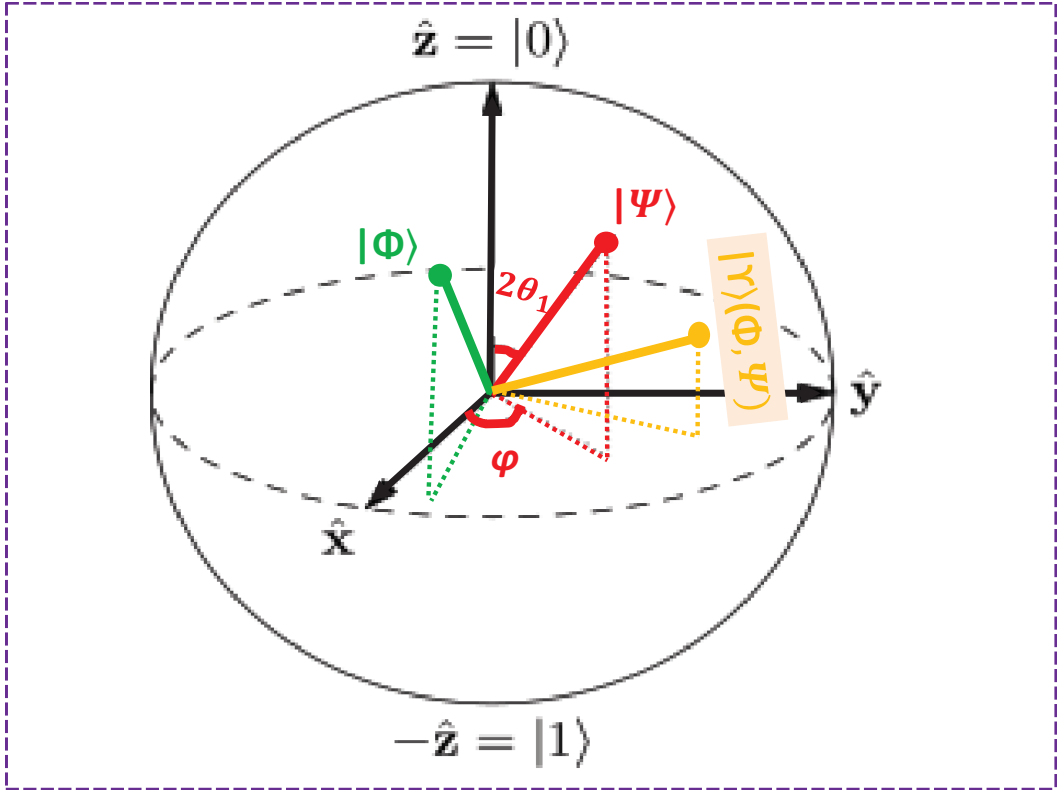}}
\caption{\footnotesize Single qubit representation on $S^{2}$}
\label{qubitrep}
\end{figure}

To a $n$-qubit quantum computer, we are interested in the synchronized $n$-qubit quantum computations used in its internal operations concerning addition (+), substraction (-), multiplication ($\ast$), and division ($\slash$), which are corresponding to those used in a conventional single-digit computer. In doing so, let $|\Phi\rangle$ and $|\Psi\rangle$ be two $n$-qubit vectors that satisfy the constraint in \eq{cnqubi} and have the expressions in \eq{sphericalco} with associated spherical coordinates
\begin{eqnarray}
\theta^{\Phi}=(\theta^{\Phi}_{1},...,\theta^{\Phi}_{2^{n}})\;\;\;\mbox{and}\;\;\;\theta^{\Psi}=(\theta^{\Psi}_{1},...,\theta^{\Psi}_{2^{n}})
\elabel{qwasqea}
\end{eqnarray}
respectively. Then, we have the following $n$-qubit quantum operational rules in terms of addition (+), substraction (-), multiplication ($\ast$), and division ($\slash$). However, all the operational results should satisfy the constraint in \eq{cnqubi}, i.e., all the related $n$-qubit quantum wave functions should keep to locate on the unit $(n+1)$-sphere (denoted by $S^{n+1}$ and see Figure~\ref{qubitrep} for an example over $S^{2}$ corresponding to $n=1$).
More precisely, we have that
\begin{eqnarray}
\left\{\begin{array}{ll}
|\Phi\rangle+|\Psi\rangle&=\;\;\sum_{j_{l}\in\{0,1\},\;l\in\{1,...,n\}}\left(\phi_{j_{1}...j_{n}}+\psi_{j_{1}...j_{n}}\right)|j_{1}...j_{n}\rangle,\\
|\Phi\rangle-|\Psi\rangle&=\;\;\sum_{j_{l}\in\{0,1\},\;l\in\{1,...,n\}}\left(\phi_{j_{1}...j_{n}}-\psi_{j_{1}...j_{n}}\right)|j_{1}...j_{n}\rangle,\\
|\Phi\rangle*|\Psi\rangle&=\;\;\sum_{j_{l}\in\{0,1\},\;l\in\{1,...,n\}}\left(\phi_{j_{1}...j_{n}}*\psi_{j_{1}...j_{n}}\right)|j_{1}...j_{n}\rangle,\\
|\Phi\rangle/|\Psi\rangle&=\;\;\sum_{j_{l}\in\{0,1\},\;l\in\{1,...,n\}}\left(\phi_{j_{1}...j_{n}}/\psi_{j_{1}...j_{n}}\right)|j_{1}...j_{n}\rangle.
\end{array}
\right.
\elabel{fsghasda}
\end{eqnarray}
Note that, the $n$-qubit quantum wave functions introduced in \eq{fsghasda} may not be on $S^{n+1}$. Therefore, to justify them meaningful in terms of addition (+), substraction (-), multiplication ($\ast$), and division ($\slash$) over $S^{n+1}$, we have the following theorem.
\begin{proposition}\label{fgshajksda}
The statement of this proposition consists of the following two parts:

\begin{enumerate}
\item (Part I.) For a $n$-qubit quantum wave function $|\Psi\rangle$ whose coefficients are given by \eq{sphericalco}, it satisfies the constraint in \eq{cnqubi}.
\item (Part II.) If $|\Phi\rangle$ and $|\Psi\rangle$ are two $n$-qubit vectors that satisfy the constraint in \eq{cnqubi} and have the expressions in \eq{sphericalco} with associated spherical coordinates in \eq{qwasqea}. Then, there are unique maps from the $n$-qubit quantum wave functions defined in \eq{fsghasda} to those $|\Upsilon\rangle(\Phi,\Psi)$ on $S^{n+1}$, i.e.,
\begin{eqnarray}
|\Upsilon\rangle(\Phi,\Psi)\in\bigg\{|\Upsilon\rangle^{\Phi+\Psi},\;|\Upsilon\rangle^{\Phi-\Psi},\;|\Upsilon\rangle^{\Phi*\Psi},
\;|\Upsilon\rangle^{\Phi/\Psi}\bigg\},
\elabel{jkdsalp}
\end{eqnarray}
such that those wave functions on $S^{n+1}$ satisfy the constraint in \eq{cnqubi} with the corresponding spherical coordinates given by
\begin{eqnarray}
\left\{\begin{array}{ll}
\theta^{\Phi+\Psi}_{\Upsilon}&=\;\;
\Big(\frac{\theta^{\Phi}_{1}+\theta^{\Psi}_{1}}{2},...,\frac{\theta^{\Phi}_{2^{n}}+\theta^{\Psi}_{2^{n}}}{2}\Big),\\ \theta^{\Phi-\Psi}_{\Upsilon}&\;=\;\;\;
\Big(\frac{\theta^{\Phi}_{1}-\theta^{\Psi}_{1}}{2},...,\frac{\theta^{\Phi}_{2^{n}}-\theta^{\Psi}_{2^{n}}}{2}\Big),\\ \theta^{\Phi*\Psi}_{\Upsilon}&\;\;=\;\;\;\;\Big(\theta^{\Phi}_{1}+\theta^{\Psi}_{1},...,\theta^{\Phi}_{2^{n}}+\theta^{\Psi}_{2^{n}}\Big),\\ \theta^{\Phi/\Psi}_{\Upsilon}&\;\;\;=\;\;\;\;\;\Big(\theta^{\Phi}_{1}-\theta^{\Psi}_{1},...,\theta^{\Phi}_{2^{n}}-\theta^{\Psi}_{2^{n}}\Big).
\end{array}
\right.
\elabel{fsghasdaII}
\end{eqnarray}
\end{enumerate}
\end{proposition}

The proof of this proposition is provided in Appendix (i.e., Section~\ref{fsgdhja}) of this paper. Instead, in the following subsection, we first give some illustration and examples concerning the usage of our newly introduced $n$-qubit operations as stated in the proposition.

\vskip 1cm
\subsection{Illustration and Examples}\label{alpdsa}

Consider two general $n$-qubit vectors $|\hat{\Phi}\rangle$ and $|\hat{\Psi}\rangle$ without the constraint in \eq{cnqubi}, such as, $|\hat{\Phi}\rangle=(0,0,...,0,0,1,1)$ and $|\hat{\Psi}\rangle=(1,0,...,0,1,0,1)$. Let $|\Phi\rangle$ and $|\Psi\rangle$ be their corresponding normalized $n$-qubit vectors that satisfy the constraint in \eq{cnqubi} and have the expressions in \eq{sphericalco} with associated spherical coordinates $\theta^{\Phi}=(\theta^{\Phi}_{1},...,\theta^{\Phi}_{2^{n}})$ and $\theta^{\Psi}=(\theta^{\Psi}_{1},...,\theta^{\Psi}_{2^{n}})$ respectively. Furthermore, their associated normalized constants are respectively denoted by $\| |\hat{\Phi}\rangle\|$ and $\| |\hat{\Psi}\rangle\|$. For examples, $\| |\hat{\Phi}\rangle\|=\sqrt{2}$ if $|\hat{\Phi}\rangle=(0,0,...,0,0,1,1)$ and $\| |\hat{\Psi}\rangle\|=\sqrt{3}$ if $|\hat{\Psi}\rangle=(1,0,...,0,1,0,1)$. In our designed $n$-qubit quantum computer, we are interested in the synchronized $n$-qubit quantum computation operations concerning addition (+), substraction (-), multiplication ($\ast$), and division ($\slash$) in a certain way. For example, if we consider the addition (+) operation between $|\hat{\Phi}\rangle$ and $|\hat{\Psi}\rangle$ (i.e., to use a synchronized $n$-qubit quantum computation method to compute $|\hat{\Phi}\rangle+|\hat{\Psi}\rangle$), the normalized constant of the summation $|\hat{\Phi}\rangle+|\hat{\Psi}\rangle$ is given by
\begin{eqnarray}
\left\| |\hat{\Phi}\rangle+|\hat{\Psi}\rangle\right\|=\sqrt{\left\| |\hat{\Phi}\rangle\right\|^{2}+\left\| |\hat{\Psi}\rangle\right\|^{2}}
\elabel{tyurqasa}
\end{eqnarray}
since all the components of $|\hat{\Phi}\rangle$ and $|\hat{\Psi}\rangle$ are nonnegative. With this known normalized constant, we can develop a device to detect and determine the $2^{n}$-dimensional angle vector (i.e., phase vector) $\theta^{\Phi+\Psi}_{\Upsilon}$ as derived in \eq{fsghasdaII} simultaneously. Then, we can map the corresponding $|\Upsilon\rangle(\Phi,\Psi)$ on $S^{n+1}$ back to the targeted vector $|\hat{\Phi}\rangle+|\hat{\Psi}\rangle$ through the obtained phase vector and the known normalized constant.

It is worth to point out that, by combining the multi-input multi-output (MIMO) orbit angular momentum (OAM) and random phase techniques (see, e.g., Dai~\cite{dai:optrat}), a quantum channel model is designed in Dai~\cite{dai:quacom}. This channel model can be used to interact with the recently developed Jiuzhang computer core (see, e.g., Figure~\ref{quantumchannel}).
\begin{figure}[tbh]
\centerline{\epsfxsize=6.0in\epsfbox{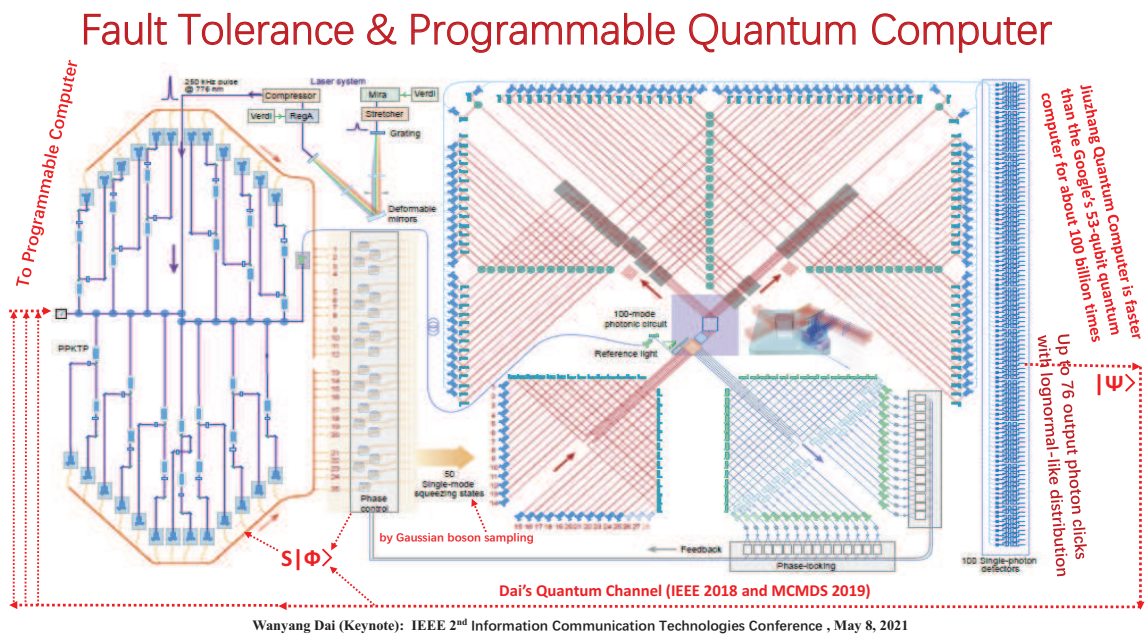}}
\caption{\footnotesize An example of the interaction between the quantum channel model in Dai~\cite{dai:quacom} and a photon based quantum computer core, where the inner photon detecting and measurement part is adapted from Zhong {\it et al.}~\cite{zho:quacom}.}
\label{quantumchannel}
\end{figure}
However, a challenging issue to the Jiuzhang computer developed in Zhong {\it et al.}~\cite{zho:quacom} is how to evolve itself to a programmable quantum computer for business usages. Therefore, due to this motivation, we first upgrade our quantum channel model designed in Dai~\cite{dai:quacom} to the one having more functionalities concerning $n$-qubit operations over $S^{n+1}$ with the target to implement a photon or other technique based programmable quantum computer.

Finally, concerning the quantum measurement channel as designed in Figure~\ref{quantumchannel}, we can set up a relationship between the original $n$-qubit $|\Phi\rangle$ and the measured $n$-qubit $|\Psi\rangle$ through the measurement formula derived in Dai~\cite{dai:quacom}, i.e.,
\begin{eqnarray}
|\Psi\rangle=|\Phi\rangle G(\Phi,\Psi)\;\;\;\mbox{with}\;\;\;G(\Phi,\Psi)={\cal H}^{\dag}(\phi)\left({\cal H}(\phi){\cal H}^{\dag}(\phi)\right)^{-1}{\cal H}(\psi)
\elabel{channelphipsi},
\end{eqnarray}
where, ${\cal H}(\phi)$ and its conjugate complex function ${\cal H}^{\dag}(\phi)$ are given by
\begin{eqnarray}
{\cal H}(\phi)=(\phi_{h_{0}},...,\phi_{h_{2^{n}-1}})',\;\;\;\;{\cal H}^{\dag}(\phi)=(\phi^{\dag}_{h_{0}},...,\phi^{\dag}_{h_{2^{n}-1}}),
\elabel{hhdag}
\end{eqnarray}
for all index $h\in\{h_{0},h_{1},...,h_{2^{n}-1}\}=\{0,1,...,2^{n}-1\}$ as introduced in \eq{reindexi}. Similarly, ${\cal H}(\psi)$ and ${\cal H}^{\dag}(\psi)$ can also be expressed in this way. However, from the original $n$-qubit $|\Phi\rangle$ to the measured $n$-qubit $|\Psi\rangle$ as displayed in Figure~\ref{quantumchannel}, it takes time and we call this time as processing time (or service time). Due to this time delay, the incoming $n$-qubit data may not be immediately processed and hence they will be stored in queueing buffers as designed in Figure~\ref{quantumchannel}. Therefore, in the subsequent discussion of this paper, we will focus on studying the queueing dynamics and conducting its associated performance modeling.

\vskip 1cm
\section{Queueing Scaling Limits}\label{queuelimit}

In this section, we study the performance modeling concerning the internal data flow dynamics through deriving the scaling limits (i.e., RGRFs on a $(n+1)$-sphere) for $n$-qubit quantum computer based queueing systems under two different heavy traffic regimes. In Subsection~\ref{hjslwse}, we introduce the required preliminaries and present our performance measures. In Subsection~\ref{mainthm}, we state our heavy traffic limit theorems. In Subsection~\ref{proofthm}, we prove these limit theorems.

\subsection{Preliminaries and Performance Measures}\label{hjslwse}

In this subsection, we suppose that the quantum computer serves $J$ queues in parallel (indexed by $j\in{\cal J}\equiv\{1,...,J\}$ and corresponding to $J$ users) as shown in Figure~\ref{quantumchannel}. Each user's data arrival stream is formed in quantum data packets. The size of each $n$-qubit data packet is supposed to be a random number $\zeta\in\{1,2,...\}$. In other words, each quantum data packet can be denoted by a sequence of $n$-qubits $\{|\Phi_{1}\rangle,...,|\Phi_{\zeta}\rangle\}$, where, $|\Phi_{i}\rangle$ for each $i\in\{1,2,...,\zeta\}$ denotes a $n$-qubit and satisfies the constraint in \eq{cnqubi}. Therefore, associated with the queues, there is a $J$-dimensional data arrival process
\begin{eqnarray}
A=\{A(t,X)=(A_{1}(t,X),...,A_{J}(t,X))',t\geq 0,X\subset S^{n+1}\},
\elabel{arrivala}
\end{eqnarray}
where, $A_{j}(t,X)$ with $j\in{\cal J}$ is the number of $n$-qubit based data packets that arrive at the $j$th queue during $(0,t]$ over a subset $X$ of $S^{n+1}$. Note that, here and elsewhere in the paper, the prime denotes the transpose of a vector or a matrix. 
Then, we can state the formal definition of the DSRRRF introduced in Introduction of this paper as follows.
\begin{definition}\label{dsrp}
A random field $A_{j}(\cdot,\cdot)$ with $j\in{\cal J}\equiv\{1,...,J\}$ on $S^{n+1}$ is called a DSRRRF over $S^{n+1}$ if $A_{j}(\cdot,X)$ is the counting process corresponding to a renewal reward process with arrival rate $\lambda_{j}(X)$ and mean reward $m_{j}$ associated with finite squared coefficients of variations $\alpha^{2}_{j}(X)$ and $\zeta^{2}_{j}$.
\end{definition}
In addition, we let $\{u_{j}(k,X),k\in\{1,2,...\}\}$ with $X\in S^{n+1}$ be the sequence of times between the arrivals of the $(k-1)$th and the $k$th reward batches of packets at the $j$th queue. The corresponding batch reward is denoted by $w_{j}(k,X)$ and all the packets arrived with it are indexed in certain successive order. Then, we can define the renewal counting process associated with the inter-arrival time sequence
$\{u_{j}(k,X),k\in\{1,2,...\}\}$ for each $j\in{\cal J}$ by
\begin{eqnarray}
N_{j}(t,X)=\sup\left\{m\geq 0:\sum_{k=1}^{m}u_{j}(k,X)\leq t\right\}.
\elabel{nsjvc}
\end{eqnarray}
Hence, we can present the DSRRP $A_{j}(\cdot,\cdot)$ via
\begin{eqnarray}
A_{j}(t,X)=\sum_{k=1}^{N_{j}(t,X)}w_{j}(k,X).
\elabel{ansjvc}
\end{eqnarray}
Each $n$-qubit based packet will first get service in the quantum computer and then leave it, where the computer is assumed to operates under a non-idling work-conserving policy (i.e., if there is any customer in the system, the computer will not stop working). Furthermore, we let $\{v_{j}(k,X),k=1,2,...\}$ with $X\in S^{n+1}$ be the sequence of successive arrived packet lengths at queue $j$, which is assumed to be a sequence of strictly positive $i.i.d.$ random variables with average packet length $1/\mu_{j}\in(0,\infty)$ and squared coefficient of variation $\beta_{j}^{2}\in(0,\infty)$. In addition, we assume that all inter-arrival and service time processes are mutually independent. For each $j\in{\cal J}$ and each nonnegative constant $h$, we use $S_{j}(\cdot,\cdot)$ to denote the renewal counting process associated with $\{v_{j}(k,X),k=1,2,...\}$, i.e.,
\begin{eqnarray}
S_{j}(h,X)=\sup\left\{m\geq 0:\sum_{k=1}^{m}v_{j}(k,X)\leq h\right\}
\elabel{sjvc}
\end{eqnarray}
and its corresponding vector form will be denoted by $S(\cdot,\cdot)$.

Let $Q_{j}(t,X)$ be the $j$th queue length with $j\in{\cal J}$ at each time $t\in[0,\infty)$
and $D_{j}(t,X)$ be the number of packet departures from the $j$th queue in $(0,t]$.
Then, the queueing dynamics governing the evolving of data in and data out in the quantum computer can be modeled by
\begin{eqnarray}
Q_{j}(t,X)=Q_{j}(0,X)+A_{j}(t,X)-D_{j}(t,X),
\elabel{queuelength}
\end{eqnarray}
where, each queue is supposed to have an infinite storage capacity to buffer $n$-qubit based data packets arrived for a given user. Furthermore, let $B_{j}(t,X)$ denote the cumulative amount of busy time devoted to user $j$ by $t$ and $\Lambda_{j}(X)$ be the corresponding service rate allocated to user $j$. Then, we know that
\begin{eqnarray}
D_{j}(t,X)=S_{j}(B_{j}(t,X),X).
\elabel{deprata}
\end{eqnarray}
In addition, we use $V(t,X)$ and $W_{j}(t,X)$ to denote the (expected) total workload over $X$ in the quantum computer at time $t$ and the one corresponding to user $j\in{\cal J}$ at time $t$, i.e.,
\begin{eqnarray}
V(t)=\sum_{j=1}W_{j}(t,X),\;\;\;\;\;\;\;\;\;W_{j}(t,X)=\frac{Q_{j}(t,X)}{\mu_{j}}.
\elabel{mldagsj}
\end{eqnarray}
Finally, the corresponding vector forms of the previously related random fields will be denoted by $Q(\cdot,\cdot)$, $D(\cdot,\cdot)$, $B(\cdot,\cdot)$, and $W(\cdot,\cdot)$ for later references. Note that, the total workload random field $V(\cdot,\cdot)$ will be used as our performance measure in the subsequent study of this paper.

\vskip 1cm
\subsection{Heavy Traffic Limit Theorems}\label{mainthm}

In this subsection, we present our limit theorems for our queueing systems under two different heavy traffic regimes in the following two subsections. In the first regime that corresponds to the way used in Dai~\cite{dai:broapp,dai:optrat} and Dai and Dai~\cite{daidai:healim}, the qubit number $n$ is given and fixed. In the second regime, the qubit number $n$ is considered as a varying scaling parameter, which is in contrast to the well-known Halfin-Whitt regime in~\cite{halwhi:heatra} where the number of servers is considered as a scaling parameter.

\vskip 1cm
\subsubsection{The Case Corresponding to Fixed Qubit Number}~\label{qosplaswq}

In this subsubsection, we consider the case that the qubit number $n$ is given and fixed in our quantum computer based server. Therefore, in this case, we are interested in conducting the performance modeling in terms of the relationship between input data rates and service rates. In doing so, for a point $x\in S^{n+1}$, let
\begin{eqnarray}
\Big\{X^{k}:X^{k}\subset S^{n+1},k\in{\cal R}\Big\}\;\;\;\mbox{with}\;\;\;{\cal R}\equiv\{1,2,...\}
\nonumber
\end{eqnarray}
be a decreasing subset sequence of $S^{n}$ such that
\begin{eqnarray}
\bigcap_{k=1}^{\infty}X^{k}=X=\{x\}.
\nonumber
\end{eqnarray}
An example with $n=1$ concerning the relationship between $x$ and $X^{k}$ is shown in the green and brown circular areas over $S^{2}$ in Figure~\ref{qubitrandomwalk}. Furthermore, we use $d^{k}=|X^{k}|$ to denote the surface area size of $X^{k}$. Then, we can define two sequences of diffusion-scaled processes $\{\hat{V}^{rk}(\cdot,\cdot):\;r,k\in{\cal R}\}$ and $\{\hat{Q}^{rk}(\cdot,\cdot):\;r,k\in{\cal R}\}$ corresponding to a sequence of arrival rates $\{\lambda^{rk}(X^{k})=(\lambda_{1}^{rk}(X^{k}),...,\lambda^{rk}_{J}(X^{k}))':\;r,k\in{\cal R}\}$ and a sequence of service rates $\{\Lambda^{rk}(X^{k})=(\Lambda_{1}^{rk}(X^{k}),...,\Lambda_{J}^{rk}(X^{k}))':\;r,k\in{\cal R}\}$, i.e.,
\begin{eqnarray}
\hat{V}^{rk}(t,X^{k})\equiv\frac{1}{\sqrt{r}}V^{rk}(rt,X^{k}),
\;\;\;\;\;
\hat{Q}_{j}^{rk}(t,X^{k})\equiv\frac{1}{\sqrt{r}}Q_{j}^{rk}(rt,X^{k})
\elabel{wklque}
\end{eqnarray}
for each $t\geq 0$.  In addition, let
\begin{eqnarray}
\mu^{rk}(X^{k})&=&\sum_{j=1}^{J}\frac{1}{\mu_{j}}\left(m_{j}\lambda^{rk}_{j}(X^{k})-\Lambda^{rk}_{j}(X^{k})\right).
\elabel{htraf}
\end{eqnarray}
Then, we can impose the following so-called heavy traffic condition in terms of both time and area-size scalings by evolving the way used in Dai~\cite{dai:broapp,dai:optrat} and Dai and Dai~\cite{daidai:healim},
\begin{eqnarray}
\left\{\begin{array}{ll}
\lambda^{rk}_{j}(X^{k})\;\rightarrow\;\lambda^{k}_{j}(X^{k})    &\mbox{as}\;\;r\rightarrow\infty\;\;\mbox{for a fixed}\;\;k\in{\cal R},\\
\alpha^{rk}_{j}(X^{k})\;\rightarrow\;\alpha^{k}_{j}(X^{k})      &\mbox{as}\;\;r\rightarrow\infty\;\;\mbox{for a fixed}\;\;k\in{\cal R},\\
\Lambda^{rk}_{j}(X^{k})\;\rightarrow\;\Lambda^{k}_{j}(X^{k})    &\mbox{as}\;\;r\rightarrow\infty\;\;\mbox{for a fixed}\;\;k\in{\cal R},\\
\sqrt{r}\mu^{rk}(X^{k})\;\rightarrow\;\theta^{k}(X^{k}) &\mbox{as}\;\;r\rightarrow\infty\;\;\mbox{for a fixed}\;\;k\in{\cal R},\\
\theta^{k}(X^{k})\;\rightarrow\;\theta(x)             &\mbox{as}\;\;k\rightarrow\infty,\\
\lambda^{k}_{j}(X^{k})\;\rightarrow\;\lambda_{j}(x)   &\mbox{as}\;\;k\rightarrow\infty,\\
\alpha^{k}_{j}(X^{k})\;\rightarrow\;\alpha_{j}(x)     &\mbox{as}\;\;k\rightarrow\infty,\\
\Lambda^{k}_{j}(X^{k})\;\rightarrow\;\Lambda_{j}(x)   &\mbox{as}\;\;k\rightarrow\infty,\\
X^{k}\;\rightarrow\;x                                 &\mbox{as}\;\;k\rightarrow\infty,\\
d^{k}\;\rightarrow\;0                                 &\mbox{as}\;\;k\rightarrow\infty,
\end{array}
\right.
\elabel{heaconI}
\end{eqnarray}
where, the limit $\theta^{k}(X^{k})$ for each given $k\in{\cal R}$ is a constant, and the further limit $\theta$(x) is also a constant for each fixed $x\in S^{n+1}$ (but not depending on $k$). Furthermore, we suppose that
\begin{eqnarray}
(\hat{Q}^{rk}_{1}(0,X^{k}),...,\hat{Q}^{rk}_{J}(0,X^{k}))&\Rightarrow&(\tilde{Q}^{k}_{1}(0,X^{k}),...,\tilde{Q}^{k}_{J}(0,X^{k}))\;\;\;
\mbox{as}\;\; r\rightarrow\infty\;\;\mbox{for a $k$},
\elabel{fdghjds}\\
(\tilde{Q}^{k}_{1}(0,X^{k}),...,\tilde{Q}^{k}_{J}(0,X^{k}))&\Rightarrow&(\tilde{Q}_{1}(0,x),...,\tilde{Q}_{J}(0,x))\;\;\;\;\;\;\;\;
\mbox{as}\;\;k\rightarrow\infty,
\elabel{fdghjdsI}
\end{eqnarray}
where, the notation ``$\Rightarrow$" denotes the convergence in distribution and $\tilde{Q}_{j}(0,x)$ is a Gaussian random variable at $x\in S^{n+1}$. Then, we can present our first main theorem as follows.
\begin{theorem}\label{thmone}
Under the heavy traffic condition presented in \eq{heaconI}, the following convergence in distribution is true, i.e.,
\begin{eqnarray}
\hat{V}^{rk}(\cdot,\cdot)\Rightarrow\hat{V}(\cdot,\cdot)
\elabel{aI-qwyweakc}
\end{eqnarray}
as $r\rightarrow\infty$ first and $k\rightarrow\infty$ second, where $\hat{V}(\cdot,\cdot)$ is a RGRF over ${\cal M}=S^{n+1}$ as stated in Definition~\ref{defrgrf} with $\mu(t,x)=\theta t$.
\end{theorem}

\vskip 1cm
\subsubsection{The Case Corresponding to Variable Qubit Number}

In this subsubsection, we consider the case that the qubit number $n$ is a varying scaling parameter, which is in contrast to the well-known Halfin-Whitt regime. In this case, besides conducting the performance modeling in terms of the relationship between input data rates and service rates, we also need to study their relationship with the qubit number $n$ with the aim to determine a suitable parameter $n$ corresponding to given input and service rates. In doing so, the scaling parameter $k$ in Subsection~\ref{qosplaswq} is taken to be the qubit number $n$ (i.e., $k=n$). Furthermore, for a point $x\in S^{\infty}$, let $\{X^{n}:\;X^{n}\subset S^{n+1},n\in{\cal R}\}$ be a subset sequence such that $X^{n}\rightarrow x$ as $n\rightarrow\infty$. Then, we can present our second main theorem as follows.
\begin{theorem}\label{thmtwo}
Under the heavy traffic condition presented in \eq{heaconI}, the following convergence in distribution is true, i.e.,
\begin{eqnarray}
\hat{V}^{rn}(\cdot,\cdot)\Rightarrow\hat{V}(\cdot,\cdot)
\elabel{hw-qwyweakc}
\end{eqnarray}
as $r\rightarrow\infty$ first and $n\rightarrow\infty$ second, where $\hat{V}(\cdot,\cdot)$ is a RGRF over ${\cal M}=S^{\infty}$ as stated in Definition~\ref{defrgrf} with $\mu(t,x)=\theta t$.
\end{theorem}



\vskip 1cm
\subsection{Proofs of Heavy Traffic Limit Theorems}\label{proofthm}

First of all, we remark that the proof of Theorem~\ref{thmtwo} is essentially the same as the one of Theorem~\ref{thmone}. Thus, we here only provide a proof for Theorem~\ref{thmone}. In doing so, it follows from the expressions in \eq{queuelength}, \eq{mldagsj}, and \eq{wklque} that
\begin{eqnarray}
\hat{V}^{rk}(t,X^{k})=\hat{V}^{rk}(0,X^{k})+\hat{U}^{rk}(t,X^{k})+\sqrt{r}\mu^{rk}(X^{k})t+\sqrt{r}\Lambda^{rk}(X^{k})\bar{B}^{rk}(t,X^{k}),
\elabel{waseqr}
\end{eqnarray}
where, $\Lambda^{rk}=(\Lambda^{rk}_{1}(X^{k})/\mu_{1}\;,...,\Lambda^{rk}_{J}(X^{k})/\mu_{J})$, and
\begin{eqnarray}
\hat{V}^{rk}(0,X^{k})&=&\sum_{j=1}^{J}\frac{1}{\mu_{j}}\hat{Q}^{rk}_{j}(0,X^{k}),
\nonumber\\
\hat{Q}^{rk}_{j}(0,X^{k})&=&\frac{1}{\sqrt{r}}Q^{rk}_{j}(0,X^{k}),
\nonumber\\
\hat{U}^{rk}(t,X^{k})&=&\sum_{j=1}^{J}\frac{1}{\mu_{j}}\hat{U}^{rk}_{j}(t,X^{k}),
\nonumber\\
\hat{U}^{rk}_{j}(t,X^{k})&=&\hat{A}^{rk}_{j}(rt,X^{k})-\hat{S}^{rk}_{j}(\bar{B}^{rk}_{j}(t,X^{k}),X^{k}),
\nonumber\\
\hat{A}^{rk}_{j}(t,X^{k})&=&\frac{1}{\sqrt{r}}\bigg(A^{rk}_{j}(rt,X^{k})-\left(rm_{j}\lambda^{rk}_{j}(X^{k})\right)t\bigg),
\nonumber\\
\hat{S}^{rk}_{j}(t,X^{k})&=&\frac{1}{\sqrt{r}}\bigg(S^{rk}_{j}(rt,X^{k})-\left(r\Lambda^{rk}_{j}(X^{k})\right)t\bigg),
\nonumber\\
\bar{B}^{rk}(t,X^{k})&=&\bigg(\bar{B}^{rk}_{1}(t,X^{k})\;,...,\bar{B}^{rk}_{J}(t,X^{k})\bigg)',
\nonumber\\
\bar{B}^{rk}_{j}(t,X^{k})&=&\frac{1}{r}B^{rk}_{j}(rt,X^{k}).
\nonumber
\end{eqnarray}
Let $e_{i}=(0,...,0,1,0,...,0)'$ denote the $J$-dimensional unit vector whose $i$th component is the unity and others are all zero. Then, it follows from \eq{waseqr} that
\begin{eqnarray}
e_{i}\hat{V}^{rk}(t,X^{k})&=&e_{i}\hat{V}^{rk}(0,X^{k})+e_{i}\hat{U}^{rk}(t,X^{k})
\elabel{aawaseqr}\\
&&+e_{i}\sqrt{r}\mu^{rk}(X^{k})t+\sqrt{r}e_{i}\Lambda^{rk}(X^{k})\bar{B}^{rk}(t,X^{k}).
\nonumber
\end{eqnarray}
Furthermore, let $\Xi^{r}$ be a constant given by $\Xi^{r}=(e'(E\Lambda^{r})^{-1}Ee)^{-1}$, where, $E=\mbox{diag}(1,...,1)$ is the $J\times J$ diagonal matrix with the unity in the main diagonal and $e$ is the $1\times J$ matrix given by $e=(1,...,1)'$. Then, it follows from \eq{aawaseqr} that
\begin{eqnarray}
\hat{V}^{rk}(t,X^{k})=\hat{V}^{rk}(0,X^{k})+\hat{U}^{rk}(t,X^{k})+\sqrt{r}\mu^{rk}(X^{k})t+\hat{I}^{rk}(t,X^{k})
\elabel{newwaseqr}
\end{eqnarray}
with
\begin{eqnarray}
\hat{I}^{rk}(t,X^{k})=\Xi^{r}\sqrt{r}\left(t-\sum_{j=1}^{J}\bar{B}^{rk}_{j}(t,X^{k})\right).
\elabel{newwaseqrI}
\end{eqnarray}
Note that, for each $r,k\in{\cal R}$, due to the non-idling work-conserving policy, we know that
\begin{eqnarray}
\int_{0}^{\infty}\hat{V}^{rk}(t,X^{k})d\hat{I}^{rk}(t,X^{k})=0.
\elabel{aargrfp2}
\end{eqnarray}
In addition, similar to the vector expression for $\bar{B}^{rk}$, we use $\hat{A}^{rk}$, $\hat{Q}^{rk}$, $\hat{V}^{rk}$, $\hat{U}^{rk}$, and $\hat{S}^{rk}$ to denote the vector counterparts of the concerned processes. Then, due to the functional central limit theorem for triply stochastic renewal reward processes (see the discussion in Dai~\cite{dai:plamod}) or simply for renewal reward processes (see, e.g., Whitt~\cite{whi:stopro}), we know that
\begin{eqnarray}
\hat{A}^{rk}(t,X^{k})\Rightarrow\tilde{A}^{k}(t,X^{k})\;\;\;\mbox{and}\;\;\;\hat{S}^{rk}(t,X^{k})\Rightarrow\tilde{S}^{k}(t,X^{k})\;\;\;\mbox{as}\;\;\; r\rightarrow\infty
\elabel{jklsaqwa}
\end{eqnarray}
for each fixed $k\in{\cal R}$, where the limits $\tilde{A}^{k}$ and $\tilde{S}^{k}$ are Brownian motions depending on the area $X^{k}\subset S^{n}$ through their associated arrival and service rates. More precisely, the means of $\tilde{A}^{k}$ and $\tilde{S}^{k}$ are both zero. Their covariance matrices for the fixed $k$ are respectively given by
\begin{eqnarray}
\Gamma^{A^{k}}&=&\mbox{diag}\bigg(m^{2}_{1}\left(\lambda_{1}^{k}(X^{k})\zeta_{1}^{2}+\lambda_{1}(X^{k})\alpha_{1}^{2}(X^{k})\right),...,
\elabel{gammaa}\\
&&\;\;\;\;\;\;\;\;\;\;\;\;\;\;\;\;\;\;\;\;\;\;\;\;\;\;\;\;\;\;\;\;\;\;\;\;
m_{J}^{2}\left(\lambda_{J}^{k}(X^{k})\zeta_{J}^{2}+\lambda_{J}(X^{k})\alpha_{J}^{2}(X^{k})\right)\bigg),
\nonumber\\
\Gamma^{S^{k}}&=&\mbox{diag}\bigg(\Lambda^{k}_{1}(X^{k})\beta_{1}^{2},...,\Lambda^{k}_{J}(X^{k})\beta_{J}^{2}\bigg),
\elabel{gammas}
\end{eqnarray}
where, ``diag" means diagonal matrix. Moreover, by \eq{jklsaqwa}-\eq{gammas}, we can conclude that
\begin{eqnarray}
\tilde{A}^{k}(t,X^{k})\Rightarrow\tilde{A}(t,x)\;\;\;\mbox{and}\;\;\;\tilde{S}^{k}(t,X^{k})\Rightarrow\tilde{S}(t,x)\;\;\;\mbox{as}\;\;\; k\rightarrow\infty,
\elabel{jklsaqwaI}
\end{eqnarray}
where, the limits $\tilde{A}$ and $\tilde{S}$ are Brownian motions depending on $x\in S^{n+1}$ through their associated arrival and service rates. More precisely, the means of $\tilde{A}$ and $\tilde{S}$ are both zero. Their covariance matrices are respectively given by
\begin{eqnarray}
\Gamma^{A}&=&\mbox{diag}\bigg(m^{2}_{1}\Big(\lambda_{1}(x)\zeta_{1}^{2}
+\lambda_{1}(x)\alpha_{1}^{2}(x)\Big),...,m_{J}^{2}\Big(\lambda_{J}(x)\zeta_{J}^{2}+\lambda_{J}(x)\alpha_{J}^{2}(x)\Big)\bigg),
\elabel{gammaaI}\\
\Gamma^{S}&=&\mbox{diag}\bigg(\Lambda_{1}(x)\beta_{1}^{2},...,\Lambda_{J}(x)\beta_{J}^{2}\bigg).
\elabel{gammasI}
\end{eqnarray}
To further conduct our proof, we define the associated fluid scaling process $\bar{F}^{rk}$ for each functional $\hat{F}^{rk}\in\{\hat{Q}^{rk}$, $\hat{V}^{rk}$, $\hat{U}^{r}$, $\hat{S}^{r}$, $\hat{I}^{rk}\}$ by
\begin{eqnarray}
\bar{F}^{rk}(t,X^{k})=\frac{1}{\sqrt{r}}\hat{F}^{rk}(rt,X^{k}).
\elabel{plfsaq}
\end{eqnarray}
Then, it follows from \eq{newwaseqr} that
\begin{eqnarray}
\bar{V}^{rk}(t,X^{k})=\bar{V}^{rk}(0,X^{k})+\bar{U}^{rk}(t,X^{k})+\mu^{rk}(X^{k})t+\bar{I}^{rk}(t,X^{k}).
\elabel{barnewwaseqr}
\end{eqnarray}
Hence, by the property in \eq{aargrfp2}, it follows from \eq{barnewwaseqr} and the Skorohod mapping (see, e.g., Dai~\cite{dai:broapp,dai:optrat} and Dai and Dai~\cite{daidai:healim}) that
\begin{eqnarray}
\bar{I}^{rk}(t,X^{k})=\displaystyle\sup_{0\leq s \leq t}\Big(\bar{V}^{rk}(0,X^{k})+\bar{U}^{rk}(s,X^{k})+\mu^{rk}(X^{k})s\Big)^{-},
\elabel{rgrfp}
\end{eqnarray}
where, for a real number $a$, $(a)^{-}=-a$ if $a\leq 0$ and $(a)^{-}=0$ otherwise. Note that, since
\begin{eqnarray}
\Big|\bar{B}^{rk}_{j}(t_{2},X^{k})-\bar{B}^{rk}_{j}(t_{1},X^{k})\Big|\leq \Big|t_{2}-t_{1}\Big|
\elabel{lipschitzc}
\end{eqnarray}
for any $t_{1},t_{2}\in[0,\infty)$ and each $j\in{\cal J}$, $\bar{B}^{rk}_{j}(t,X^{k})$ is Lipschitz continuous. Thus, it follows from \eq{newwaseqrI} that $\bar{I}^{rk}(t,X^{k})$ is also Lipschitz continuous. Therefore, by the heavy traffic condition in \eq{heaconI}, the weak convergence in \eq{fdghjds} and \eq{jklsaqwa}, and the property in \eq{lipschitzc}, it follows from the Skorohod representation theorem (see, e.g., Ethier and Kurtz\cite{ethkur:marpro}) and the discussion in Dai~\cite{dai:broapp,dai:optrat} and Dai and Dai~\cite{daidai:healim}) that
\begin{eqnarray}
\bar{I}^{rk}(t,X^{k})\rightarrow 0\;\;\;\;\mbox{u.o.c.}\;\;\;\mbox{a.s.}\;\;\;\mbox{as}\;\;\;r\rightarrow\infty\;\;\;\mbox{for each given}\;\;\; k\in{\cal R},
\elabel{icong}
\end{eqnarray}
where, {\it u.o.c.} means ``uniform convergence on compact set of $[0,\infty)$". Hence, we know that
\begin{eqnarray}
\bar{B}^{rk}_{j}(t,X^{k})\rightarrow t\;\;\;\;\mbox{u.o.c.}\;\;\;\mbox{a.s.}\;\;\;\mbox{as}\;\;\;r\rightarrow\infty\;\;\;\mbox{for each given}\;\;\; k\in{\cal R}.
\elabel{icongI}
\end{eqnarray}
Then, by \eq{fdghjds}-\eq{fdghjdsI}, \eq{heaconI}, \eq{jklsaqwa}-\eq{gammas}, \eq{icong}-\eq{icongI}, the continuous mapping theorem (see, e.g., Billingsley~\cite{bil:conpro}), the Skorohod representation theorem (see, e.g., Ethier and Kurtz\cite{ethkur:marpro}), and the way developed in Dai~\cite{dai:broapp,dai:optrat} and Dai and Dai~\cite{daidai:healim}), we know that
\begin{eqnarray}
\hat{V}^{rk}(0,X^{k})+\hat{U}^{rk}(t,X^{k})+\sqrt{r}\mu^{rk}(X^{k})t\rightarrow\tilde{V}^{k}(0,X^{k})+\tilde{U}^{k}(t,X^{k})+\theta^{k}t
\elabel{netpcon}
\end{eqnarray}
u.o.c. a.s. as $r\rightarrow\infty$ for each given $k\in{\cal R}$, where,
\begin{eqnarray}
\tilde{V}^{k}(0,X^{k})&=&\sum_{j=1}^{J}\frac{1}{\mu_{j}}\tilde{Q}^{k}_{j}(0,X^{k}),
\nonumber\\
\tilde{U}^{k}(t,X^{k})&=&\tilde{A}^{k}(t,X^{k})-\tilde{S}^{k}(t,X^{k}).
\nonumber
\end{eqnarray}
Furthermore, by \eq{netpcon} and the continuous mapping theorem, we know that
\begin{eqnarray}
\hat{I}^{rk}(t,X^{k})\rightarrow\tilde{I}^{k}(t,X^{k})=\displaystyle\sup_{0\leq s \leq t}\Big(\tilde{V}^{k}(0,X^{k})+\tilde{U}^{k}(s,X^{k})+\theta^{k}(X^{k})s\Big)^{-}
\elabel{abcdsad}
\end{eqnarray}
u.o.c. a.s. as $r\rightarrow\infty$ for each given $k\in{\cal R}$, which implies that
\begin{eqnarray}
\hat{V}^{rk}(t,X^{k})\rightarrow\tilde{V}^{k}(t,X^{k})=\tilde{V}^{k}(0,X^{k})+\tilde{U}^{k}(t,X^{k})+\theta^{k}(X^{k})t+\tilde{I}^{k}(t,X^{k}).
\elabel{abnewwaseqr}
\end{eqnarray}
In addition, by \eq{aargrfp2}, \eq{abcdsad}, and the continuous mapping theorem, we have that
\begin{eqnarray}
0=\int_{0}^{\infty}\hat{V}^{rk}(t,X^{k})d\hat{I}^{rk}(t,X^{k})\rightarrow\int_{0}^{\infty}\tilde{V}^{k}(t,X^{k})d\tilde{I}^{k}(t,X^{k})
\elabel{aargrfp2I}
\end{eqnarray}
u.o.c. a.s. as $r\rightarrow\infty$ for each given $k\in{\cal R}$.

Hence, it follows from the similar discussion in proving the claims in \eq{abnewwaseqr}-\eq{aargrfp2I} that
\begin{eqnarray}
\tilde{I}^{k}(t,X^{k})&\rightarrow&\tilde{I}(t,x)\;=\;\displaystyle\sup_{0\leq s \leq t}\Big(\tilde{V}(0,x)+\tilde{U}(s,x)+\theta(x)s\Big)^{-},
\elabel{sabcdsad}\\
\tilde{V}^{k}(t,X^{k})&\rightarrow&\tilde{V}(t,x)=\tilde{V}(0,x)+\tilde{U}(t,x)+\theta(x)t+\tilde{I}(t,x).
\elabel{sabnewwaseqr}
\end{eqnarray}
u.o.c. a.s. as $k\rightarrow\infty$. Moreover, we have that
\begin{eqnarray}
0&=&\int_{0}^{\infty}\hat{V}^{k}(t,X^{k})d\hat{I}^{k}(t,X^{k})\rightarrow\int_{0}^{\infty}\tilde{V}(t,x)d\tilde{I}(t,x)
\elabel{saargrfp2I}
\end{eqnarray}
u.o.c. a.s. as $k\rightarrow\infty$.

Finally, by applying the Skorohod representation theorem back to \eq{sabcdsad}-\eq{saargrfp2I}, we know that the claim in Theorem~\ref{thmone} is true. $\Box$

\vskip 1cm
\section{Appendix: Proof of Proposition~\ref{fgshajksda}}~\label{fsgdhja}

The proof of Proposition~\ref{fgshajksda} consists of the following two parts that correspond to the two counterparts in the statement of the theorem.

\subsection{Proof of Part I.}

In this part of proof, we prove the claim that a $n$-qubit quantum wave function $|\Psi\rangle$ satisfies the constraint in \eq{cnqubi} if its coefficients are given by \eq{sphericalco}. In fact, for the example shown in Figure~\ref{qubitrep} with $n=1$, the single-qubit quantum wave function $|\Psi\rangle=\psi_{0}|0\rangle+\psi_{1}|1\rangle$ with $\psi_{0}=\mbox{cos}\left(\theta_{1}\right)$ and
$\psi_{1}=e^{i\varphi}\mbox{sin}(\theta_{1})$ and satisfies $|\psi_{0}|^{2}+|\psi_{1}|^{2}=1$. Furthermore, if $n=2$, the coefficients of the corresponding $2$-qubit quantum wave function in \eq{nqubi}, which is given by
\begin{eqnarray}
|\Psi\rangle=\psi_{0}|00\rangle+\psi_{1}|01\rangle+\psi_{2}|10\rangle+\psi_{3}|11\rangle,
\nonumber
\end{eqnarray}
also satisfy the relationship in \eq{cnqubi} due to the following computation,
\begin{eqnarray}
&&\left|\psi_{0}\right|^{2}+\left|\psi_{1}\right|^{2}+\left|\psi_{2}\right|^{2}+\left|\psi_{3}\right|^{2}
\nonumber\\
&=&\left|\cos(\theta_{1})\right|^{2}+\left|\sin(\theta_{1})\cos(\theta_{2})\right|^{2}
+\left|\sin(\theta_{1})\sin(\theta_{2})\cos(\theta_{3})\right|^{2}
+\left|e^{i\varphi}\sin(\theta_{1})\sin(\theta_{2})\sin(\theta_{3})\right|^{2}
\nonumber\\
&=&\left|\cos(\theta_{1})\right|^{2}+\left|\sin(\theta_{1})\right|^{2}\left|\cos(\theta_{2})\right|^{2}
+\left|\sin(\theta_{1})\right|^{2}\left|\sin(\theta_{2})\right|^{2}\left(|\cos(\theta_{3})|^{2}+|\sin(\theta_{3})|^{2}\right)
\nonumber\\
&=&\left|\cos(\theta_{1})\right|^{2}+\left|\sin(\theta_{1})\right|^{2}\left(|\cos(\theta_{2})|^{2}+|\sin(\theta_{2})|^{2}\right)
\nonumber\\
&=&1.
\nonumber
\end{eqnarray}
By the similar way, we can show that the constraint in \eq{cnqubi} is true for the given expressions in \eq{sphericalco} corresponding to each $n\in\{1,2,...\}$.

\subsection{Proof of Part II}

In this part of proof, we prove the claims concerning the $n$-qubit quantum operations to be true. The proof consists of four steps corresponding to the four operations of addition (+), substraction (-), multiplication ($\ast$), and division ($\slash$) over $S^{n+1}$.

\vskip 0.3cm 
{\bf Step I.}$\;\;$ For the addition (+) operation, the corresponding coefficients of $|\Upsilon\rangle(\Phi,\Psi)=|\Upsilon\rangle^{\Phi+\Psi}$ in \eq{jkdsalp} can be calculated as follows. For $k=1$, we have that
\begin{eqnarray}
\upsilon_{1}(\phi_{1},\psi_{1})=\frac{\phi_{1}+\psi_{1}}{2\cos\left(\frac{\theta^{\Phi}_{1}-\theta^{\Psi}_{1}}{2}\right)}
=\frac{\cos\left(\theta^{\Phi}_{1}\right)
+\cos\left(\theta^{\Psi}_{1}\right)}{2\cos\left(\frac{\theta^{\Phi}_{1}-\theta^{\Psi}_{1}}{2}\right)}
=\cos\left(\frac{\theta^{\Phi}_{1}+\theta^{\Psi}_{1}}{2}\right).
\elabel{lsadass}
\end{eqnarray}
For $k=2$, we have that
\begin{eqnarray}
\upsilon_{2}(\phi_{2},\psi_{2})&=&\frac{(\phi_{2}+\psi_{2})+\sin(\theta^{\Phi}_{1})\cos(\theta^{\Psi}_{2})
+\sin(\theta^{\Psi}_{1})\cos(\theta^{\Phi}_{2})}{2^{2}\cos\left(\frac{\theta^{\Phi}_{1}-\theta^{\Psi}_{1}}{2}\right)
\cos\left(\frac{\theta^{\Phi}_{2}-\theta^{\Psi}_{2}}{2}\right)}
\elabel{lsadassI}\\
&=&\frac{\left(\sin(\theta^{\Phi}_{1})+\sin(\theta^{\Psi}_{1})\right)\left(\cos(\theta^{\Phi}_{2})+\cos(\theta^{\Psi}_{2})\right)}
{2^{2}\cos\left(\frac{\theta^{\Phi}_{1}-\theta^{\Psi}_{1}}{2}\right)\cos\left(\frac{\theta^{\Phi}_{2}-\theta^{\Psi}_{2}}{2}\right)}
\nonumber\\
&=&\sin\left(\frac{\theta^{\Phi}_{1}+\theta^{\Psi}_{1}}{2}\right)\cos\left(\frac{\theta^{\Phi}_{1}+\theta^{\Psi}_{1}}{2}\right).
\nonumber
\end{eqnarray}
In general, for an integer $k\in\{2,3,...,2^{n}-1\}$, we have that
\begin{eqnarray}
\upsilon_{k}(\phi_{k},\psi_{k})&=&\frac{1}{2^{k}\prod_{j=1}^{k}\cos\left(\frac{\theta^{\Phi}_{j}-\theta^{\Psi}_{j}}{2}\right)}
\elabel{lsadassII}\\
&&\;\;\;\;\;\;\;
\bigg((\phi_{k}+\psi_{k})+\Big(\prod_{j=1}^{k-1}\left(\sin(\theta^{\Phi}_{j})+\sin(\theta^{\Psi}_{j})\right)
\left(\cos(\theta^{\Phi}_{k})+\cos(\theta^{\Psi}_{k})\right)\Big)
\nonumber\\
&&\;\;\;\;\;\;\;\;\;\;\;\;\;\;\;\;\;\;\;\;\;\;\;\;\;\;\;\;\;\;\;\;
-\Big(\prod_{j=1}^{k-1}\sin(\theta^{\Phi}_{j})\cos(\theta^{\Phi}_{k})+\prod_{j=1}^{k-1}\sin(\theta^{\Psi}_{j})\cos(\theta^{\Psi}_{k})\Big)\bigg)
\nonumber\\
&=&\frac{\prod_{j=1}^{k-1}\left(\sin(\theta^{\Phi}_{j})+\sin(\theta^{\Psi}_{j})\right)\left(\cos(\theta^{\Phi}_{k})+\cos(\theta^{\Psi}_{k})\right)}
{2^{k}\prod_{j=1}^{k}\cos\left(\frac{\theta^{\Phi}_{j}-\theta^{\Psi}_{j}}{2}\right)}
\nonumber\\
&=&\prod_{j=1}^{k-1}\sin\left(\frac{\theta^{\Phi}_{j}+\theta^{\Psi}_{j}}{2}\right)\cos\left(\frac{\theta^{\Phi}_{k}+\theta^{\Psi}_{k}}{2}\right).
\nonumber
\end{eqnarray}
Finally, for $k=2^{n}$, we have that
\begin{eqnarray}
\upsilon_{2^{n}}(\phi_{2^{n}},\psi_{2^{n}})&=&\frac{1}{2^{2^{n-1}}e^{\left(i\left(\theta_{2^{n}}^{\Phi}+\theta_{2^{n}}^{\Psi}\right)/2\right)}
\prod_{j=1}^{2^{n}-1}\cos\left(\frac{\theta^{\Phi}_{j}-\theta^{\Psi}_{j}}{2}\right)}
\elabel{lsadassIII}\\
&&\;\;\;\;\;\;\;\;\;\;
\bigg((\phi_{2^{n}}+\psi_{2^{n}})+e^{i\left(\theta_{2^{n}}^{\Phi}+\theta_{2^{n}}^{\Psi}\right)}
\prod_{j=1}^{2^{n}-1}\left(\sin(\theta^{\Phi}_{j})+\sin(\theta^{\Psi}_{j})\right)
\nonumber\\
&&\;\;\;\;\;\;\;\;\;\;\;\;\;\;\;\;\;\;\;\;\;\;\;\;\;\;\;\;\;\;\;\;
-\Big(e^{i\theta_{2^{n}}^{\Phi}}\prod_{j=1}^{2^{n}-1}\sin(\theta^{\Phi}_{j})
+e^{i\theta_{2^{n}}^{\Psi}}\prod_{j=1}^{2^{n}-1}\sin(\theta^{\Psi}_{j})\Big)\bigg)
\nonumber\\
&=&\frac{e^{i\left(\theta_{2^{n}}^{\Phi}+\theta_{2^{n}}^{\Psi}\right)}
\prod_{j=1}^{2^{n}-1}\left(\sin(\theta^{\Phi}_{j})+\sin(\theta^{\Psi}_{j})\right)}
{2^{2^{n-1}}e^{\left(i\left(\theta_{2^{n}}^{\Phi}+\theta_{2^{n}}^{\Psi}\right)/2\right)}
\prod_{j=1}^{2^{n}-1}\cos\left(\frac{\theta^{\Phi}_{j}-\theta^{\Psi}_{j}}{2}\right)}
\nonumber\\
&=&e^{\left(i\left(\theta_{2^{n}}^{\Phi}+\theta_{2^{n}}^{\Psi}\right)/2\right)}
\prod_{j=1}^{2^{n}-1}\sin\left(\frac{\theta^{\Phi}_{j}+\theta^{\Psi}_{j}}{2}\right).
\nonumber
\end{eqnarray}
Thus, $|\Upsilon\rangle(\Phi,\Psi)=|\Upsilon\rangle^{\Phi+\Psi}$ in \eq{jkdsalp} with the coefficients in \eq{lsadass}-\eq{lsadassIII} satisfies the constraint in \eq{cnqubi} and has the spherical coordinate $\theta^{\Phi+\Psi}_{\Upsilon}$ as in \eq{fsghasdaII}.

\vskip 0.3cm 
{\bf Step II.}$\;\;$ For the substraction (-) operation, the corresponding coefficients of $|\Upsilon\rangle(\Phi,\Psi)=|\Upsilon\rangle^{\Phi-\Psi}$ in \eq{jkdsalp} can be calculated as follows. For $k=1$, we have that
\begin{eqnarray}
\upsilon_{1}(\phi_{1},\psi_{1})=\frac{\phi_{1}-\psi_{1}+2\cos\left(\theta^{\Psi}_{1}\right)}{2\cos\left(\frac{\theta^{\Phi}_{1}+\theta^{\Psi}_{1}}{2}\right)}
=\frac{\cos\left(\theta^{\Phi}_{1}\right)+\cos\left(\theta^{\Psi}_{1}\right)}{2\cos\left(\frac{\theta^{\Phi}_{1}+\theta^{\Psi}_{1}}{2}\right)}
=\cos\left(\frac{\theta^{\Phi}_{1}-\theta^{\Psi}_{1}}{2}\right).
\elabel{mlsadass}
\end{eqnarray}
For $k=2$, we have that
\begin{eqnarray}
\upsilon_{2}(\phi_{2},\psi_{2})&=&\frac{(\phi_{2}-\psi_{2})+\sin(\theta^{\Phi}_{1})\cos(\theta^{\Psi}_{2})
-\sin(\theta^{\Psi}_{1})\cos(\theta^{\Phi}_{2})}{2^{2}\cos\left(\frac{\theta^{\Phi}_{1}-\theta^{\Psi}_{1}}{2}\right)
\cos\left(\frac{\theta^{\Phi}_{2}+\theta^{\Psi}_{2}}{2}\right)}
\elabel{mlsadassI}\\
&=&\frac{\left(\sin(\theta^{\Phi}_{1})-\sin(\theta^{\Psi}_{1})\right)\left(\cos(\theta^{\Phi}_{2})+\cos(\theta^{\Psi}_{2})\right)}
{2^{2}\cos\left(\frac{\theta^{\Phi}_{1}-\theta^{\Psi}_{1}}{2}\right)\cos\left(\frac{\theta^{\Phi}_{2}+\theta^{\Psi}_{2}}{2}\right)}
\nonumber\\
&=&\sin\left(\frac{\theta^{\Phi}_{1}-\theta^{\Psi}_{1}}{2}\right)\cos\left(\frac{\theta^{\Phi}_{1}-\theta^{\Psi}_{1}}{2}\right).
\nonumber
\end{eqnarray}
In general, for an integer $k\in\{2,3,...,2^{n}-1\}$, we have that
\begin{eqnarray}
\upsilon_{k}(\phi_{k},\psi_{k})&=&\frac{1}{2^{k}\prod_{j=1}^{k-1}\cos\left(\frac{\theta^{\Phi}_{j}-\theta^{\Psi}_{j}}{2}\right)
\cos\left(\frac{\theta^{\Phi}_{k}+\theta^{\Psi}_{k}}{2}\right)}
\elabel{mlsadassII}\\
&&\;\;\;\;\;\;\;
\bigg((\phi_{k}-\psi_{k})+\Big(\prod_{j=1}^{k-1}\left(\sin(\theta^{\Phi}_{j})-\sin(\theta^{\Psi}_{j})\right)
\left(\cos(\theta^{\Phi}_{k})+\cos(\theta^{\Psi}_{k})\right)\Big)
\nonumber\\
&&\;\;\;\;\;\;\;\;\;\;\;\;\;\;\;\;\;\;\;\;\;\;\;\;\;\;\;\;\;\;\;\;\;\;\;\;
-\Big(\prod_{j=1}^{k-1}\sin(\theta^{\Phi}_{j})\cos(\theta^{\Phi}_{k})-\prod_{j=1}^{k-1}\sin(\theta^{\Psi}_{j})\cos(\theta^{\Psi}_{k})\Big)\bigg)
\nonumber\\
&=&\frac{\prod_{j=1}^{k-1}\left(\sin(\theta^{\Phi}_{j})-\sin(\theta^{\Psi}_{j})\right)\left(\cos(\theta^{\Phi}_{k})+\cos(\theta^{\Psi}_{k})\right)}
{2^{k}\prod_{j=1}^{k-1}\cos\left(\frac{\theta^{\Phi}_{j}-\theta^{\Psi}_{j}}{2}\right)\cos\left(\frac{\theta^{\Phi}_{k}+\theta^{\Psi}_{k}}{2}\right)}
\nonumber\\
&=&\prod_{j=1}^{k-1}\sin\left(\frac{\theta^{\Phi}_{j}-\theta^{\Psi}_{j}}{2}\right)\cos\left(\frac{\theta^{\Phi}_{k}-\theta^{\Psi}_{k}}{2}\right).
\nonumber
\end{eqnarray}
Finally, for $k=2^{n}$, we have that
\begin{eqnarray}
\upsilon_{2^{n}}(\phi_{2^{n}},\psi_{2^{n}})&=&\frac{1}{2^{2^{n-1}}e^{\left(i\left(\theta_{2^{n}}^{\Phi}-\theta_{2^{n}}^{\Psi}\right)/2\right)}
\prod_{j=1}^{2^{n}-1}\cos\left(\frac{\theta^{\Phi}_{j}-\theta^{\Psi}_{j}}{2}\right)}
\elabel{mlsadassIII}\\
&&\;\;\;\;\;\;
\bigg((\phi_{2^{n}}+\psi_{2^{n}})+e^{i\left(\theta_{2^{n}}^{\Phi}-\theta_{2^{n}}^{\Psi}\right)}
\prod_{j=1}^{2^{n}-1}\left(\sin(\theta^{\Phi}_{j})-\sin(\theta^{\Psi}_{j})\right)
\nonumber\\
&&\;\;\;\;\;\;\;\;\;\;\;\;\;\;\;\;\;\;\;\;\;\;\;\;\;\;\;\;\;\;
-\Big(e^{i\theta_{2^{n}}^{\Phi}}\prod_{j=1}^{2^{n}-1}\sin(\theta^{\Phi}_{j})
-e^{i\theta_{2^{n}}^{\Psi}}\prod_{j=1}^{2^{n}-1}\sin(\theta^{\Psi}_{j})\Big)\bigg)
\nonumber\\
&=&\frac{e^{i\left(\theta_{2^{n}}^{\Phi}-\theta_{2^{n}}^{\Psi}\right)}
\prod_{j=1}^{2^{n}-1}\left(\sin(\theta^{\Phi}_{j})-\sin(\theta^{\Psi}_{j})\right)}
{2^{2^{n-1}}e^{\left(i\left(\theta_{2^{n}}^{\Phi}-\theta_{2^{n}}^{\Psi}\right)/2\right)}
\prod_{j=1}^{2^{n}-1}\cos\left(\frac{\theta^{\Phi}_{j}-\theta^{\Psi}_{j}}{2}\right)}
\nonumber\\
&=&e^{\left(i\left(\theta_{2^{n}}^{\Phi}-\theta_{2^{n}}^{\Psi}\right)/2\right)}
\prod_{j=1}^{2^{n}-1}\sin\left(\frac{\theta^{\Phi}_{j}-\theta^{\Psi}_{j}}{2}\right).
\nonumber
\end{eqnarray}
Thus, $|\Upsilon\rangle(\Phi,\Psi)=|\Upsilon\rangle^{\Phi-\Psi}$ in \eq{jkdsalp} with the coefficients in \eq{mlsadass}-\eq{mlsadassIII} satisfies the constraint in \eq{cnqubi} and has the spherical coordinate $\theta^{\Phi-\Psi}_{\Upsilon}$ as in \eq{fsghasdaII}.

\vskip 0.3cm 
{\bf Step III.}$\;\;$ For the multiplication ($\ast$) operation, the corresponding coefficients of $|\Upsilon\rangle(\Phi,\Psi)=|\Upsilon\rangle^{\Phi*\Psi}$ in \eq{jkdsalp} can be calculated as follows. For $k=1$, we have that
\begin{eqnarray}
\upsilon_{1}(\phi_{1},\psi_{1})=2\phi_{1}*\psi_{1}-\cos\left(\theta^{\Phi}_{1}-\theta^{\Psi}_{1}\right)
=\cos\left(\theta^{\Phi}_{1}+\theta^{\Psi}_{1}\right).
\elabel{smlsadass}
\end{eqnarray}
For $k=2$, we have that
\begin{eqnarray}
\upsilon_{2}(\phi_{2},\psi_{2})&=&\frac{2^{2}\cos(\theta^{\Phi}_{1})*\phi_{2}*\psi_{2}}{\sin(\theta^{\Phi}_{1})}
-\Big(\sin(\theta^{\Phi}_{1}+\theta^{\Psi}_{1})\cos(\theta^{\Phi}_{2}-\theta^{\Psi}_{2})
\elabel{smlsadassI}\\
&&\;\;\;\;\;\;\;\;\;\;\;\;\;
-\sin(\theta^{\Phi}_{1}-\theta^{\Psi}_{1})\cos(\theta^{\Phi}_{2}-\theta^{\Psi}_{2})
-\sin(\theta^{\Phi}_{1}-\theta^{\Psi}_{1})\cos(\theta^{\Phi}_{2}+\theta^{\Psi}_{2})\Big)
\nonumber\\
&=&\Big(\sin(\theta^{\Phi}_{1}+\theta^{\Psi}_{1})-\sin(\theta^{\Phi}_{1}-\theta^{\Psi}_{1})\Big)
\Big(\cos(\theta^{\Phi}_{2}-\theta^{\Psi}_{2})+\cos(\theta^{\Phi}_{2}+\theta^{\Psi}_{2})\Big)
\nonumber\\
&&\;\;\;\;\;\;\;\;\;\;\;\;\;
-\Big(\sin(\theta^{\Phi}_{1}+\theta^{\Psi}_{1})\cos(\theta^{\Phi}_{2}-\theta^{\Psi}_{2})
-\sin(\theta^{\Phi}_{1}-\theta^{\Psi}_{1})\cos(\theta^{\Phi}_{2}-\theta^{\Psi}_{2})
\nonumber\\
&&\;\;\;\;\;\;\;\;\;\;\;\;\;\;\;\;\;\;\;\;\;\;\;\;\;\;\;\;\;\;\;\;\;\;\;\;\;\;\;\;\;\;\;\;\;\;\;\;\;\;\;\;\;\;\;\;\;\;\;\;\;\;\;
-\sin(\theta^{\Phi}_{1}-\theta^{\Psi}_{1})\cos(\theta^{\Phi}_{2}+\theta^{\Psi}_{2})\Big)
\nonumber\\
&=&\sin(\theta^{\Phi}_{1}+\theta^{\Psi}_{1})\cos(\theta^{\Phi}_{2}+\theta^{\Psi}_{2}).
\nonumber
\end{eqnarray}
In general, for an integer $k\in\{2,3,...,2^{n}-1\}$, we have that
\begin{eqnarray}
\upsilon_{k}(\phi_{k},\psi_{k})&=&\frac{2^{k}\left(\prod_{j=1}^{k-1}\cos(\theta^{\Phi}_{j})\right)*\phi_{k}*\psi_{k}}
{\prod_{j=1}^{k-1}\sin(\theta^{\Phi}_{j})}
\elabel{smlsadassII}\\
&&\;\;
-\Bigg(\prod_{j=1}^{k-1}\Big(\sin(\theta^{\Phi}_{j}+\theta^{\Psi}_{j})-\sin(\theta^{\Phi}_{j}-\theta^{\Psi}_{j})\Big)
\Big(\cos(\theta^{\Phi}_{k}-\theta^{\Psi}_{k})+\cos(\theta^{\Phi}_{k}+\theta^{\Psi}_{k})\Big)
\nonumber\\
&&\;\;\;\;\;\;\;\;\;\;\;\;\;\;\;\;\;\;\;\;\;\;\;\;\;\;\;\;\;\;\;\;\;\;\;\;\;\;\;\;\;\;\;\;\;\;\;\;\;\;\;\;\;\;\;\;\;
-\prod_{j=1}^{k-1}\sin(\theta^{\Phi}_{j}+\theta^{\Psi}_{j})\cos(\theta^{\Phi}_{k}+\theta^{\Psi}_{k})\Big)\Bigg)
\nonumber\\
&=&\prod_{j=1}^{k-1}\Big(\sin(\theta^{\Phi}_{j}+\theta^{\Psi}_{j})-\sin(\theta^{\Phi}_{j}-\theta^{\Psi}_{j})\Big)
\Big(\cos(\theta^{\Phi}_{k}-\theta^{\Psi}_{k})+\cos(\theta^{\Phi}_{k}+\theta^{\Psi}_{k})\Big)
\nonumber\\
&&\;\;
-\Bigg(\prod_{j=1}^{k-1}\Big(\sin(\theta^{\Phi}_{j}+\theta^{\Psi}_{j})-\sin(\theta^{\Phi}_{j}-\theta^{\Psi}_{j})\Big)
\Big(\cos(\theta^{\Phi}_{k}-\theta^{\Psi}_{k})+\cos(\theta^{\Phi}_{k}+\theta^{\Psi}_{k})\Big)
\nonumber\\
&&\;\;\;\;\;\;\;\;\;\;\;\;\;\;\;\;\;\;\;\;\;\;\;\;\;\;\;\;\;\;\;\;\;\;\;\;\;\;\;\;\;\;\;\;\;\;\;\;\;\;\;\;\;\;\;\;\;\;\;\;
-\prod_{j=1}^{k-1}\sin(\theta^{\Phi}_{j}+\theta^{\Psi}_{j})\cos(\theta^{\Phi}_{k}+\theta^{\Psi}_{k})\Big)\Bigg)
\nonumber\\
&=&\prod_{j=1}^{k-1}\sin(\theta^{\Phi}_{j}+\theta^{\Psi}_{j})\cos(\theta^{\Phi}_{k}+\theta^{\Psi}_{k}).
\nonumber
\end{eqnarray}
Finally, for $k=2^{n}$, we have that
\begin{eqnarray}
\upsilon_{2^{n}}(\phi_{2^{n}},\psi_{2^{n}})&=&\frac{2^{2^{n}-1}
\prod_{j=1}^{2^{n}-1}\cos(\theta^{\Phi}_{j})*\phi_{2^{n}}*\psi_{2^{n}}}{\prod_{j=1}^{2^{n}-1}\sin(\theta^{\Phi}_{j})}
\elabel{smlsadassIII}\\
&&\;\;\;\;\;\;\;\;\;
-e^{i\left((\theta_{2^{n}}^{\Phi}+\theta_{2^{n}}^{\Psi})\right)}
\Bigg(\prod_{j=1}^{2^{n}-1}\Big(\sin(\theta^{\Phi}_{j}+\theta^{\Psi}_{j})-\sin(\theta^{\Phi}_{j}-\theta^{\Psi}_{j})\Big)
\nonumber\\
&&\;\;\;\;\;\;\;\;\;\;\;\;\;\;\;\;\;\;\;\;\;\;\;\;\;\;\;\;\;\;\;\;\;\;\;\;\;\;\;\;\;\;\;\;\;\;\;\;\;\;\;\;\;\;\;\;\;\;\;\;\;\;\;\;\;\;\;\;
-\prod_{j=1}^{2^{n}-1}\sin(\theta^{\Phi}_{j}+\theta^{\Psi}_{j})\bigg)\Bigg)
\nonumber\\
&=&e^{i\left((\theta_{2^{n}}^{\Phi}+\theta_{2^{n}}^{\Psi})\right)}
\Bigg(\prod_{j=1}^{2^{n}-1}\Big(\sin(\theta^{\Phi}_{j}+\theta^{\Psi}_{j})-\sin(\theta^{\Phi}_{j}-\theta^{\Psi}_{j})\Big)
\nonumber\\
&&\;\;\;\;\;
-\bigg(\prod_{j=1}^{2^{n}-1}\Big(\sin(\theta^{\Phi}_{j}+\theta^{\Psi}_{j})-\sin(\theta^{\Phi}_{j}-\theta^{\Psi}_{j})\Big)
-\prod_{j=1}^{2^{n}-1}\sin(\theta^{\Phi}_{j}+\theta^{\Psi}_{j})\bigg)\Bigg)
\nonumber\\
&=&e^{\left(i(\theta_{2^{n}}^{\Phi}+\theta_{2^{n}}^{\Psi})\right)}
\prod_{j=1}^{2^{n}-1}\sin\left(\theta^{\Phi}_{j}+\theta^{\Psi}_{j}\right).
\nonumber
\end{eqnarray}
Thus, $|\Upsilon\rangle(\Phi,\Psi)=|\Upsilon\rangle^{\Phi*\Psi}$ in \eq{jkdsalp} with the coefficients in \eq{smlsadass}-\eq{smlsadassIII} satisfies the constraint in \eq{cnqubi} and has the spherical coordinate $\theta^{\Phi*\Psi}_{\Upsilon}$ as in \eq{fsghasdaII}.

\vskip 0.3cm 
{\bf Step IV.}$\;\;$ For the division ($\slash$) operation, the corresponding coefficients of $|\Upsilon\rangle(\Phi,\Psi)=|\Upsilon\rangle^{\Phi/\Psi}$ in \eq{jkdsalp} can be calculated as follows. For $k=1$, we have that
\begin{eqnarray}
\upsilon_{1}(\phi_{1},\psi_{1})=2(\phi_{1}/\psi_{1})\cos^{2}(\theta^{\Psi}_{1})-\cos\left(\theta^{\Phi}_{1}+\theta^{\Psi}_{1}\right)
=\cos\left(\theta^{\Phi}_{1}-\theta^{\Psi}_{1}\right).
\elabel{dsmlsadass}
\end{eqnarray}
For $k=2$, we have that
\begin{eqnarray}
\upsilon_{2}(\phi_{2},\psi_{2})&=&-\frac{2^{2}\cos(\theta^{\Phi}_{1})*(\phi_{2}/\psi_{2})*\sin^{2}(\theta_{1}^{\Psi})\cos^{2}(\theta^{\Psi}_{2})}
{\sin(\theta^{\Phi}_{1})}
\elabel{dsmlsadassI}\\
&&\;\;\;\;\;\;\;\;\;\;\;\;
+\Big(\sin(\theta^{\Phi}_{1}+\theta^{\Psi}_{1})\cos(\theta^{\Phi}_{2}-\theta^{\Psi}_{2})
+\sin(\theta^{\Phi}_{1}+\theta^{\Psi}_{1})\cos(\theta^{\Phi}_{2}+\theta^{\Psi}_{2})
\nonumber\\
&&\;\;\;\;\;\;\;\;\;\;\;\;\;\;\;\;\;\;\;\;\;\;\;\;\;\;\;\;\;\;\;\;\;\;\;\;\;\;\;\;\;\;\;\;\;\;\;\;\;\;\;\;\;\;\;\;\;\;\;\;\;\;\;\;\;\;\;\;
-\sin(\theta^{\Phi}_{1}-\theta^{\Psi}_{1})\cos(\theta^{\Phi}_{2}-\theta^{\Psi}_{2})\Big)
\nonumber\\
&=&-\Big(\sin(\theta^{\Phi}_{1}+\theta^{\Psi}_{1})-\sin(\theta^{\Phi}_{1}-\theta^{\Psi}_{1})\Big)
\Big(\cos(\theta^{\Phi}_{2}-\theta^{\Psi}_{2})+\cos(\theta^{\Phi}_{2}+\theta^{\Psi}_{2})\Big)
\nonumber\\
&&\;\;\;\;\;\;\;\;\;\;\;\;
+\Big(\sin(\theta^{\Phi}_{1}+\theta^{\Psi}_{1})\cos(\theta^{\Phi}_{2}-\theta^{\Psi}_{2})
+\sin(\theta^{\Phi}_{1}+\theta^{\Psi}_{1})\cos(\theta^{\Phi}_{2}+\theta^{\Psi}_{2})
\nonumber\\
&&\;\;\;\;\;\;\;\;\;\;\;\;\;\;\;\;\;\;\;\;\;\;\;\;\;\;\;\;\;\;\;\;\;\;\;\;\;\;\;\;\;\;\;\;\;\;\;\;\;\;\;\;\;\;\;\;\;\;\;\;\;\;\;\;\;\;\;\;
-\sin(\theta^{\Phi}_{1}-\theta^{\Psi}_{1})\cos(\theta^{\Phi}_{2}-\theta^{\Psi}_{2})\Big)
\nonumber\\
&=&\sin(\theta^{\Phi}_{1}-\theta^{\Psi}_{1})\cos(\theta^{\Phi}_{2}-\theta^{\Psi}_{2}).
\nonumber
\end{eqnarray}
In general, for an integer $k\in\{2,3,...,2^{n}-1\}$, we have that
\begin{eqnarray}
\upsilon_{k}(\phi_{k},\psi_{k})&=&-\frac{(-1)^{k-1}2^{k}\left(\prod_{j=1}^{k-1}\cos(\theta^{\Phi}_{j})\right)*(\phi_{k}/\psi_{k})
*\prod_{j=1}^{k-1}\sin^{2}(\theta^{\Psi}_{j})\cos^{2}(\theta^{\Psi}_{k})}
{\prod_{j=1}^{k-1}\sin(\theta^{\Phi}_{j})}
\elabel{dsmlsadassII}\\
&&
-\Bigg(\prod_{j=1}^{k-1}\Big(-\left(\sin(\theta^{\Phi}_{j}+\theta^{\Psi}_{j})-\sin(\theta^{\Phi}_{j}-\theta^{\Psi}_{j})\right)\Big)
\Big(\cos(\theta^{\Phi}_{k}-\theta^{\Psi}_{k})+\cos(\theta^{\Phi}_{k}+\theta^{\Psi}_{k})\Big)
\nonumber\\
&&\;\;\;\;\;\;\;\;\;\;\;\;\;\;\;\;\;\;\;\;\;\;\;\;\;\;\;\;\;\;\;\;\;\;\;\;\;\;\;\;\;\;\;\;\;\;\;\;\;\;\;\;\;\;\;\;\;\;\;\;\;\;\;\;\;\;
-\prod_{j=1}^{k-1}\sin(\theta^{\Phi}_{j}-\theta^{\Psi}_{j})\cos(\theta^{\Phi}_{k}-\theta^{\Psi}_{k})\Big)\Bigg)
\nonumber\\
&=&\prod_{j=1}^{k-1}\Big(-\left(\sin(\theta^{\Phi}_{j}+\theta^{\Psi}_{j})-\sin(\theta^{\Phi}_{j}-\theta^{\Psi}_{j})\right)\Big)
\Big(\cos(\theta^{\Phi}_{k}-\theta^{\Psi}_{k})+\cos(\theta^{\Phi}_{k}+\theta^{\Psi}_{k})\Big)
\nonumber\\
&&
-\Bigg(\prod_{j=1}^{k-1}\Big(-\left(\sin(\theta^{\Phi}_{j}+\theta^{\Psi}_{j})-\sin(\theta^{\Phi}_{j}-\theta^{\Psi}_{j})\right)\Big)
\Big(\cos(\theta^{\Phi}_{k}-\theta^{\Psi}_{k})+\cos(\theta^{\Phi}_{k}+\theta^{\Psi}_{k})\Big)
\nonumber\\
&&\;\;\;\;\;\;\;\;\;\;\;\;\;\;\;\;\;\;\;\;\;\;\;\;\;\;\;\;\;\;\;\;\;\;\;\;\;\;\;\;\;\;\;\;\;\;\;\;\;\;\;\;\;\;\;\;\;\;\;\;\;\;\;\;\;\;
-\prod_{j=1}^{k-1}\sin(\theta^{\Phi}_{j}-\theta^{\Psi}_{j})\cos(\theta^{\Phi}_{k}-\theta^{\Psi}_{k})\Big)\Bigg)
\nonumber\\
&=&\prod_{j=1}^{k-1}\sin(\theta^{\Phi}_{j}-\theta^{\Psi}_{j})\cos(\theta^{\Phi}_{k}-\theta^{\Psi}_{k}).
\nonumber
\end{eqnarray}
Finally, for $k=2^{n}$, we have that
\begin{eqnarray}
\upsilon_{2^{n}}(\phi_{2^{n}},\psi_{2^{n}})&=&\frac{(-2)^{2^{n}-1}
\prod_{j=1}^{2^{n}-1}\cos(\theta^{\Phi}_{j})*\phi_{2^{n}}*\psi_{2^{n}}*\prod_{j=1}^{2^{n}-1}\sin^{2}(\theta^{\Psi}_{j})\cos^{2}(\theta^{\Psi}_{2^{n}})}
{\prod_{j=1}^{2^{n}-1}\sin(\theta^{\Phi}_{j})}
\elabel{dsmlsadassIII}\\
&&\;\;\;\;\;\;\;\;\;\;\;\;
-e^{i\left((\theta_{2^{n}}^{\Phi}-\theta_{2^{n}}^{\Psi})\right)}
\Bigg(\prod_{j=1}^{2^{n}-1}\Big(-\left(\sin(\theta^{\Phi}_{j}+\theta^{\Psi}_{j})-\sin(\theta^{\Phi}_{j}-\theta^{\Psi}_{j})\right)\Big)
\nonumber\\
&&\;\;\;\;\;\;\;\;\;\;\;\;\;\;\;\;\;\;\;\;\;\;\;\;\;\;\;\;\;\;\;\;\;\;\;\;\;\;\;\;\;\;\;\;\;\;\;\;\;\;\;\;\;\;\;\;\;\;\;\;\;\;\;\;\;\;\;\;\;\;\;\;
-\prod_{j=1}^{2^{n}-1}\sin(\theta^{\Phi}_{j}-\theta^{\Psi}_{j})\bigg)\Bigg)
\nonumber\\
&=&e^{i\left((\theta_{2^{n}}^{\Phi}-\theta_{2^{n}}^{\Psi})\right)}
\Bigg(\prod_{j=1}^{2^{n}-1}\Big(-\left(\sin(\theta^{\Phi}_{j}+\theta^{\Psi}_{j})-\sin(\theta^{\Phi}_{j}-\theta^{\Psi}_{j})\right)\Big)
\nonumber\\
&&\;\;
-\prod_{j=1}^{2^{n}-1}\Big(-\left(\sin(\theta^{\Phi}_{j}+\theta^{\Psi}_{j})-\sin(\theta^{\Phi}_{j}-\theta^{\Psi}_{j})\right)\Big)
+\prod_{j=1}^{2^{n}-1}\sin(\theta^{\Phi}_{j}-\theta^{\Psi}_{j})\Bigg)
\nonumber\\
&=&e^{\left(i(\theta_{2^{n}}^{\Phi}-\theta_{2^{n}}^{\Psi})\right)}
\prod_{j=1}^{2^{n}-1}\sin\left(\theta^{\Phi}_{j}-\theta^{\Psi}_{j}\right).
\nonumber
\end{eqnarray}
Thus, $|\Upsilon\rangle(\Phi,\Psi)=|\Upsilon\rangle^{\Phi/\Psi}$ in \eq{jkdsalp} with the coefficients in \eq{dsmlsadass}-\eq{dsmlsadassIII} satisfies the constraint in \eq{cnqubi} and has the spherical coordinate $\theta^{\Phi/\Psi}_{\Upsilon}$ as in \eq{fsghasdaII}.

\vskip 0.3cm
Finally, by following from the proofs in Part I and Part II with four proving steps, we can reach a proof for Proposition~\ref{fgshajksda}. $\Box$

\section{Conclusion}\label{cremark}

In this paper, we study $n$-qubit operation rules on $(n+1)$-sphere with the target to help developing a (photon or other technique) based programmable quantum computer. In the meanwhile, we derive the scaling limits (referred to as RGRFs on the sphere $S^{n+1}$) for $n$-qubit quantum computer based queueing systems under two different heavy traffic regimes. The queueing systems are with multiple classes of users and batch quantum random walks over the sphere as arrival inputs. In the first regime, the qubit number $n$ is fixed and the scaling is in terms of both time and space. Under this regime, performance modeling during deriving the scaling limit RGRF in terms of reasonably balancing the arrival and service rates under first-in first-out and work conserving service policy is conducted. In the second regime, besides the time and space scaling parameters, the qubit number $n$ itself is also considered as a varying scaling parameter with the additional aim to find a suitable number of qubits for the design of a quantum computer. The second heavy traffic regime is in contrast to the well-known Halfin-Whitt regime where the number of servers is considered as a scaling parameter.

\bibliographystyle{nonumber}

\end{document}